\documentclass{article}

\def\giorno{January 2009}

\def\eqref#1{(\ref{#1})}

\def\.#1{\dot #1}

\def\D{{\mathcal D}}

\def\G{{\cal G}}

\def\L{{\cal L}}
\def\M{{\cal M}}

\def\P{{\cal P}}

\def\R{{\bf R}}  

\def\T{{\rm T}}

\def\X{{\cal X}}

\def\ss{\subset}
\def\pa{\partial}

\def\=#1{{\overline{#1}}}
\def\~#1{\widetilde #1}
\def\.#1{\dot #1}
\def\^#1{\widehat #1}

\def \wt#1{{\widetilde #1}}

\def\eb{{\bf e}}
\def\fb{{\bf f}}
\def\d{{\rm d}}       

\def\grad{\nabla}     

\def\({\left(}
\def\){\right)}
\def\[{\left[}
\def\]{\right]}

\def\a{\alpha}

\def\b{\beta}

\def\ga{\gamma}
\def\de{\delta}   
\def\eps{\varepsilon}
\def\phi{\varphi}
\def\la{\lambda}
\def\La{\Lambda}
\def\s{\sigma}
\def\om{\omega}
\def\vth{\vartheta}
\def\eps{\varepsilon}
\def\Om{\Omega}
\def\vphi{\varphi}
\def\Ga{\Gamma}
\def\th{\theta}

\def\mapright#1{\smash{\mathop{\longrightarrow}\limits^{#1}}}
\def\mapdown#1{\Big\downarrow\rlap{$\vcenter{\hbox{$\scriptstyle#1$}}$}}

\def\mapne#1{\smash{\mathop{\nearrow}\limits^{#1}}}
\def\mapnw#1{\smash{\mathop{\nwarrow}\limits^{#1}}}
\def\mapse#1{\smash{\mathop{\searrow}\limits^{#1}}}
\def\mapsw#1{\smash{\mathop{\swarrow}\limits^{#1}}}

\def\Pr{{\tt Pr}}

\def\interno{\hskip 2pt \vbox{\hbox{\vbox to .18
truecm{\vfill\hbox to .25 truecm
{\hfill\hfill}\vfill}\vrule}\hrule}\hskip 2 pt}

\def\EOP{~ \hfill $\diamondsuit$} 
\def\EOR{~ \hfill $\odot$}        

\def\beq{\begin{equation}}
\def\eeq{\end{equation}}

\def\salta#1{{}}

\def\Pr#1{{\tt Pr}^{({#1})}}

\def\digamma{\Upsilon}
\def\gaub{{\mathcal G}{\mathcal B}}

\def\Q0{\Theta}

\def\gaugeref{Ble,CCS,EGH,Ish,Nak,NaS}
\def\symref{Gbook,Vin,Olv1,Ste}

\def\musymref{CGM,GM}

\def\fracor#1#2{({#1}/{#2})}

\begin{document}

\title{A gauge-theoretic description of $\mu$-prolongations,
and $\mu$-symmetries of differential equations}

\author{Giuseppe Gaeta\thanks{gaeta@mat.unimi.it} \\
{\it Dipartimento di Matematica, Universita' di Milano} \\
{\it via Saldini 50, 20133 Milano (Italy) }}

\date{{\small To be published in {\it J. Geom. Phys.} Version of
\giorno}}

\maketitle

\noindent {\bf Summary.} We consider generalized (possibly
depending on fields as well as on space-time variables) gauge
transformations and gauge symmetries in the context of general --
that is, possibly non variational nor covariant -- differential
equations. In this case the relevant principal bundle admits the
first jet bundle (of the phase manifold) as an associated bundle,
at difference with standard Yang-Mills theories. We also show how
in this context the recently introduced operation of
$\mu$-prolongation of vector fields (which generalizes the
$\la$-prolongation of Muriel and Romero), and hence
$\mu$-symmetries of differential equations, arise naturally. This
is turn suggests several directions for further development.
\par\noindent
{\tt MSC: 58J70; 35A30; 58D19; 76M60}

\section{Introduction}

The analysis and use of symmetry properties of differential
equations \cite{\symref} is by now recognized as the most powerful
general method to attack nonlinear problems, and widely used not
only in Physics (where this theory was first extensively applied)
but also in Applied Mathematics and Engineering.

The original theory of Lie-point symmetries was over the years
generalized in several directions \cite{Cic04,Gbook,Vin,Olv1,Ste}.
All these make use of the fact that once we know how a vector
field acts on independent as well as on dependent variables (i.e.
fields), we also know how it acts on field derivatives; the lift
of a transformation from the extended phase manifold (space-time
variables and fields; with due account of the relevant side --
e.g. boundary -- conditions, this is the phase bundle) to its
action on field derivatives is known in the mathematical
literature as the {\it prolongation} operation.

Despite its success, the symmetry theory of differential equations
was until recently not able to cope with certain very simple
problems which could be explicitly integrated, yet seemed to have
no symmetry underlying this integrability (see e.g.
\cite{MuRom1,Olv1} and references therein).

Muriel and Romero \cite{MuRom1} were able to solve this puzzle in
analytical terms by considering a modified prolongation operation,
and thus a new kind of symmetries, for scalar ODEs and then also
for systems of these \cite{MuTMP,MuVigo} (see also \cite{Cicrho}
in this respect). These depend on the choice of a ${\mathcal
C}^\infty$ function, denoted $\la$ in their papers; referring to
this fact the new kind of symmetries are known as ${\mathcal
C}^\infty$-symmetries, or {\it $\la$-symmetries}.

The geometrical meaning of $\la$-prolongations was then clarified
in \cite{PuS}, by means of the classical theory of characteristics
of vector fields. A different geometrical characterization, in the
language of Cartan exterior differential ideals (ideals of
differential forms), was proposed in \cite{GM}.

This opened the way for generalizing $\la$-prolongations and
$\la$-symmetries to the framework of (single, or systems of) PDEs
\cite{GM}; in this case the central object is a matrix-valued
differential one-form $\mu = \La_i \d x^i$ (the $\La_i$ being
${\mathcal C}^\infty$ matrix functions satisfying the horizontal
Maurer-Cartan equation), and these are therefore called {\it
$\mu$-prolongations} and {\it $\mu$-symmetries}.

It was then realized that for each vector field $Y$ obtained as
the $\mu$-pro\-long\-a\-ti\-on of some vector field $X$, there is
a vector field $\wt{Y}$, locally (and globally under certain
conditions) gauge-equivalent to $Y$, obtained as the standard
prolongation of a vector field $\wt{X}$, locally (and globally
under certain conditions) gauge equivalent to $X$; see \cite{CGM}
for details.\footnote{This also has some interesting consequences
in the frame of variational problems: it is possible to extend
Noether theory to $\la$ and $\mu$ symmetries; see
\cite{CGnoet,MRO}.}

This result calls for a more complete geometrical understanding of
its origin, and shows that gauge transformations play a role also
out of the well-known framework of Yang-Mills theories
\cite{\gaugeref}, and actually also for {\it non-variational
problems } and for {\it non-invariant equations} (or {\it
non-covariant equations} in physical language). The first task is
to extend the formalism so to fully include the gauge variables,
not just leaving them to the role of external
parameters.\footnote{In order to avoid any confusion, we stress
that here we consider gauge transformations more general than
those considered in standard Yang-Mills theory: (1) these may
depend on the field themselves and not only on space-time
variables; (2) we allow nonlinear actions on the fields. See the
discussion later on in this paper.}

In this note we provide a formalism including gauge variables,
i.e. set the problem in an augmented bundle; and give a precise
formulation of the $\mu$-prolongation operation in terms of such
an augmented bundle and correspondingly an enlarged set of
variables.

The reader should be warned that this geometrical understanding
does {\it not } -- at the present stage -- correspond to a
substantial computational advantage; thus the geometric
construction presented here has -- at the present stage --
interest only {\it per se}, i.e. for the understanding of the
Geometry behind twisted prolongations. On the other hand, our work
also suggests how to extend the applications of $\mu$-symmetries,
and how to further generalize them; these matters will however
only be shortly mentioned in our final discussion, see section
\ref{sec:discussion}, deferring the implementation of such
suggestions to a later time.
\medskip

We will assume the reader to be familiar with basic jet-theoretic
material and with the standard theory of symmetry of differential
equations, as given e.g. in \cite{\symref}. We will freely use
standard multi-index notation; thus a multi-index $J =
(j_1,...,j_k)$, where $j_i \in {\bf N}$, will have order $|J| =
j_1 + ... + j_k$; and for the same multi-index $J$ we will have
$D_J = D_1^{j_1} ... D_k^{j_k}$.

We use the formalism of evolutionary representatives of vector
fields \cite{\symref}, which provide an action on sections of a
bundle via (generalized) vertical vector fields describing the
action of general (proper) vector fields in the bundle. A
discussion without resorting to evolutionary representatives would
be equivalent, but would require more involved computations; on
the other hand, our discussion immediately extends (with the
standard cautions \cite{\symref} and obvious modifications) to the
full class of generalized vertical vector fields.

\bigskip
\noindent {\bf Acknowledgements.} \par\noindent I am indebted to
Giampaolo Cicogna and Paola Morando for several constructive
discussions, and to Giuseppe Marmo for ongoing encouragement. I
would also like to thank Diego Catalano-Ferraioli for discussions
about his view of auxiliary variables in the $\lambda$-symmetries
formalism.

\section{Underlying geometry}

In this section we will first set some notation regarding standard
constructions, and then introduce the {\bf gauge bundles}, which
are the essential part of our construction.

Given a fiber bundle $\P$, the space of sections in it will be
denoted as $\Ga (\P)$. The algebra of vector fields in $\P$ will
be denoted as $\X (\P)$, and that of vertical (with respect to the
bundle projection) vector fields as $\X_v (\P)$.

\subsection{Bundles, differential equations, and symmetry}

When dealing with differential equations, the independent
variables will be denoted as $x \in B$; dependent variables as $u
\in U$. Here $B$ and $U$ are smooth manifolds. We will use local
coordinates $\{ x^1 , ... , x^m \}$ in $B$ and $u = \{u^1 ,... ,
u^n \}$ in $U$. We stress that all of our considerations will be
local.

We will then consider the bundle $(M,\pi,B)$ with fiber $\pi^{-1}
(x) = U$; thus its total space is $M \simeq B \times U$. The
bundle $M$ will be our {\bf phase bundle}.\footnote{Physically,
$B$ should be thought as a region of space-time, and $U$ as a
manifold (possibly a space) in which the field $\phi$ takes
values; the global structure of the bundle also carries
information on the boundary conditions the fields are subject to
at the boundary (if any) $\pa B$ of $B$.}.

Differential equations $\Delta$ of order $k$ identify a
submanifold in the total space of the Jet bundle $J^k M$, the {\it
solution manifold} $S_\Delta \ss J^k M$.

Sections of $M$ are naturally prolonged to sections of $J^k M$;
the function $u= f(x)$ is a solution to $\Delta$ if and only if
the prolongation of the corresponding section $\s_f = (x,f(x))$ in
$\Gamma (M)$ to a section in $\Gamma (J^k M)$, call it
$\s_f^{(k)}$, is a submanifold of $S_\Delta$.

Similarly, given a vector field $X$ acting in $M$, there is a
natural prolongation of $X$ to a vector field $X^{(k)}$ in $J^k
M$. The vector field $X$ is a symmetry of $\Delta$ if and only if
it maps solutions into solutions; equivalently, if $X^{(k)}:
S_\Delta \to \T S_\Delta$.

We stress that a different fibred structure is also possible for
the jet manifolds $M^{(k)} = J^k M$. These can also be seen as
bundles over $M$; we will denote the bundle maps for these
structures as $\s_k$, so we have $(J^k M , \s_k , M)$. The
compatibility between these two structures is given by $\pi_k =
\pi \circ \s_k$.

\medskip\noindent
{\bf Remark 1.} As already mentioned, one can consider -- beside
the natural prolongation operation for vector fields mentioned
above -- some ``twisted'' (or ``deformed'') prolongation
operations, known in the literature as ``$\la$-prolongation''
(when the deformation is related to a point-dependent scale
factor) or ``$\mu$-prolon\-ga\-tion'' (when the deformation is
related to a general point-dependent linear map). These were first
introduced by Muriel and Romero \cite{MuRom1}, and quite
surprisingly turn out to be ``as useful as the natural ones'' in
analyzing differential equations. The purpose of this paper is to
investigate the geometrical structures behind this seemingly
``unreasonable effectiveness of twisted prolongations''. \EOR

\subsection{Prolongation of vector fields in $J^k M$}
\label{subsec:prolong}

The prolongation of vector fields from $M$ to $J^k M$ goes
essentially through consideration of partial derivatives of
functions corresponding to a general section in $M$ and its
transformed under the action of the vector field $X$. As well
known, if $\s_f = \{ (x,u) : \ u = f(x) \}$, then the
one-parameter group generated by $X = \phi^a (x,u) (\pa / \pa u^a)
+ \xi^i (x,u) (\pa / \pa x^i) $ maps $\s_f$ into $\s_{\wt{f}}$
with $\wt{f}^a (x) = f^a (x) + \eps [ \phi^a (x,f(x)) - \xi^i
(x,f(x)) u^a_i ]$. Properties of transformation for partial
derivatives are readily derived from this expression.

In the case of a vertical vector field (including evolutionary
representatives of general vector fields) the prolongation formula
is specially simple: in multi-index notation, the vector field $X
= \eta^a (\pa / \pa u^a)$ is prolonged to $Y = \eta^a_J (\pa / \pa
u^a_J)$, with $\eta^a_J = D_J \eta^a$. In particular we have in
recursive form \beq\label{eq:standprol} \eta^a_{J,i} \ = \ D_i \,
\eta^a_J \ . \eeq

\subsection{$\mu$-prolongations in $J^k M$}
\label{subsec:muprolong}

Let us briefly recall how $\mu$-prolongations are defined,
restricting again to vertical vector fields for the sake of
simplicity (see \cite{\musymref} for the general case). We stress
that here we work in $M$ and $J^k M$, and {\it not} in the
augmented (jet) bundle to be defined below.

Vertical vector fields in $M$ are prolonged to vector fields in
$J^k M$ via a modified procedure based on a horizontal one form
$\mu$ with values in a representation of a Lie algebra $\G$
\cite{Ble,EGH,GoS,Str}. The form $\mu$ should satisfy the
horizontal Maurer-Cartan equation \beq\label{eq:hMCmu} D \mu \ + \
{1 \over 2} \, [\mu , \mu ] \ = \ 0 \ . \eeq In terms of the local
coordinates introduced above, we have \beq\label{eq:mu} \mu \ = \
\La_i (x,u,u_x) \ \d x^i \eeq with $\La_i$ some $n \times n$
matrices (these are related to the Lie algebra $\G$ of a Lie group
$G$ acting in $U$, see \cite{CGM,GM} for details on this
relation). With this notation, the horizontal Maurer-Cartan
equation (\ref{eq:hMCmu}) reads \beq\label{eq:hMCLa} D_i \, \La_j
\ - \ D_j \, \La_i \ + \ [\La_i , \La_j ] \ = \ 0 \ . \eeq

We write vertical vector fields in $M$ as $ X_0 = \eta^a (x,u,u_x)
 (\pa / \pa u^a)$, and the corresponding vertical
vector fields in $J^k M$ as $Y_0 = \eta^a_J (\pa / \pa u^a_J )$.
The $\mu$-prolongation formula for vertical vector fields is then,
in recursive form (cf. \eqref{eq:standprol} above),
\beq\label{eq:muprol}  \eta^a_{J,i} \ = \ D_i \eta^a_J \ + \
(\La_i)^a_{\ b} \, \eta^b_J \ . \eeq The condition
(\ref{eq:hMCLa}) guarantees the $\eta^a_J$ are well defined, see
\cite{CGM,GM}.

It was shown in \cite{CGM} that (locally, and possibly globally as
well) $\mu$ can always be written as $\mu = g^{-1} {\tt D} g$, and
correspondingly $\mu$-prolonged vector fields are related  to
standardly prolonged ones via a gauge transformation. We refer to
\cite{CGM} (in particular Theorem 1 in there) for details.

\section{The gauge bundles}

Let $G$ be a Lie group and $\epsilon : G \to e$ the operator
mapping the whole Lie group $G$ into its identity element. We
assume $G$ acts on $U$ via a (possibly nonlinear) representation
$T: G \times U \to U$; at a point $p \in U$ the action on $V :=
\T_p U$ (this is the relevant action when discussing how $G$ acts
on vector fields) is described by the linearization $\Psi = D T$
of $T$. This also induces an action of the Lie algebra $\G$ of $G$
in $V = \T_p U$ via the linear representation $\psi = D \Psi$.

Let $(\ell_1 , ... , \ell_r )$ be a basis of left-invariant vector
fields in $\G$; we write $L_i = \psi (\ell_i)$ for their
representation. Any element $\xi \in \G$ can be
written\footnote{With $\a^m = \langle \ell_m , \xi \rangle$, where
$\langle . , . \rangle$ is the scalar product in $\G$. The $\a^m$
are natural coordinates in $\G$, and using these $\ell_m$ is given
by $\ell_m = \pa / \pa \a^m $.} as $\xi = \a^m \ell_m$ and acts in
$V$ via the vector field $\psi (\xi) = \a^m L_m$.

\subsection{The basic gauge bundles}

We introduce a principal bundle $(P_G,\epsilon,M)$ over $M$ with
bundle map $\epsilon$, fiber $\epsilon^{-1} (p) = G$, and total
space $P_G \simeq M \times G = \wt{M}$. This will be called the
{\bf global gauge bundle}. Sections $\ga \in \Ga (P_G)$ are
described in local coordinates by $g = g (x,u)$.

The total space $P_G = M \times G$ can also be given the structure
of a fiber bundle $(\wt{M},\wt{\pi},B)$ over $B$ with projection
$\wt{\pi} = \pi \times \epsilon$. The compatibility between these
two structures is given by $\wt{\pi} = \pi \circ \eps$.

Let us consider a reference section $\varpi \in \Ga (P_G)$; in a
tubular neighborhood $\^G \simeq M \times G_0 \simeq M \times
\G_0$ of $\varpi$ (with $G_0 \simeq \G$ a neighborhood of zero in
$G$, and $\G_0$ a neighborhood of zero in $\G$) we can use local
coordinates $(x,u,\a)$, where $\a = 0$ identifies the section
$\varpi$. We will, for ease of notation and discussion, work in $M
\times \G$; it should be kept in mind that our results will be
local in $\G$ (and hence {\it a fortiori } in $G$). The projection
from $\G$ to $\{ 0 \} \in \G$ will be denoted as $\rho$.

The manifold $M \times \G$ can also be given two different fiber
bundle structures: we can consider it as a bundle $(\^M,\^\pi,B)$
over $B$ with total space $\^M = M \times \G$, projection $\^\pi =
\pi \times \rho$ and fiber $\^\pi^{-1} (x) = U \times \G$; or as a
bundle $(\gaub , \rho , M)$ over $M$ with total space $\gaub = \^M
= M \times \G$, projection $\rho$ and fiber $\rho^{-1} (p) = \G$.
The compatibility between these two structures is given by $\^\pi
= \pi \circ \rho$.

We will denote $\^M = (M \times \G,\^\pi,B)$ as the {\bf
augmented phase bundle}, and $\gaub = (M \times \G , \rho , M)$ as
the {\bf local gauge bundle}. In the following we will also call
these the {\bf augmented bundle} and the {\bf gauge bundle} for
short.

\medskip\noindent
{\bf Remark 2.} We stress that $M$ is {\it not } an associated
bundle for this principal fiber bundle, contrary to what happens
in Yang-Mills theories (where $g=g(x)$ and does not depend on
$u$); this is the reason for some features which could appear odd
to readers familiar with standard Yang-Mills theory. On the other
hand, $(J^1 M,\s_1,M)$, {\it is } an associated bundle for
$(P_G,\epsilon,M)$, which acts on fibers $\epsilon^{-1} (m)$ via
the representation $\psi = (D \Psi)$. \EOR

\subsection{Higher order gauge bundles}

We will also introduce gauge bundles associated to higher jet
spaces $J^k M$. Thus we consider the order $k$ augmented jet
bundle ${\^M}^{(k)} = (J^k M \times \G,\^\pi_k,B)$ with fiber
$\^\pi_k^{-1} (x) = U^{(k)} \times \G$; and the order $k$ jet
gauge bundle $J^k \gaub = (J^k M \times \G , \rho_k , J^k M)$ with
fiber $\rho_k^{-1} = \G$. (Similarly, higher order bundles with
total space $\wt{M}^{(k)} \simeq J^k M \times G$ could be defined.
We will not enter into such details.)

Moreover, recall that $J^k M$ can be seen as a bundle over $M$
with projection $\s_k$, and correspondingly $J^k \^M$ can be seen
as a bundle over $\^M$ with projection $\^\s_k$.

The situation is summarized in the following diagram, which also
embodies the different double fibrations considered above.
\beq\label{diag:star} \matrix{J^k \^M & \mapright{\rho_k} & J^k M
\cr
 & & \cr
\mapdown{\^\s_k} & \matrix{\mapse{\^\pi_k} & & \mapsw{\pi_k} \cr &
B & \cr \mapne{\^\pi} & & \mapnw{\pi} \cr} & \mapdown{\s_k} \cr
 & & \cr
\^M & \mapright{\rho} & M \cr}  \eeq

\subsection{Total derivative operators in gauge jet bundles}
\label{sec:totder}

As well known, the prolongation operation is usually performed by
applying the total derivative operators in $J^k M$; see section
\ref{subsec:prolong} above. As the gauge jet bundles we are
considering have a peculiar structure (that is, they are order $k$
jet bundles for what concerns the $u$ variables, not for what
concerns the $\a$ ones), we should discuss what are the total
derivative operators to be considered in this case.

Jet spaces are equipped with a contact structure ${\mathcal C}$,
defined by the contact forms $\chi^a_J := \d u^a_J - u^a_{J,i} \d
x^i$; note that there are no contact forms associated to gauge
variables, as we are not considering jets of these. The total
derivative operators $D_i$ can then be defined in geometric terms
as the vector fields (with a component $\pa / \pa x^i$)
annihilating all the contact forms in ${\mathcal C}$; it is
immediate to check that this requirement yields just the usual
total derivative operators (associated to the $u$ variables alone)
$D_i = (\pa / \pa x^i) + u^a_{J,i} (\pa / \pa u^a_J )$. We stress
that one should {\it not } add also components ``along the gauge
variables'', i.e. of the form $\a^m_i (\pa / \pa \a^m)$.

Thus, {\it the prolongation operation leading from $\^M$ to $J^k
\^M$ should be based on the usual total derivative operators}
$D_i$, and hence does {\it not} involve derivation with respect to
the gauge variables.\footnote{It may be worth stressing, just to
avoid any possible misunderstanding, that albeit a vector field in
$\wt{M}$ (respectively, in $\^M$) will have components both in the
$M$ and in the $G$ (respectively, $\G$) directions, the
prolongation operation should be applied {\it only } to the $M$
components, as obvious from the definition of $J^k \wt{M}$ and
$J^k \^M$ above.}

This fact shows that a substantial difference exists between the
gauge bundle and the bundle obtained by simply adding new
dependent variables $\a^m$.

\section{Prolongation of vector fields in $\wt M$ and in $\^M$}
\label{sec:prolongation}

Given a vector field in $M$, this is naturally {\it prolonged }
(or lifted) to a vector field in $J^k M$. The same applies for
vector fields in $\wt{M}$ and $\^M$, which are naturally prolonged
to vector fields respectively in $J^k \wt{M}$ and $J^k \^M$.

We will denote by $\Pr{k} [\P]$ the operator of prolongation of
vector fields in a bundle $\P$ to vector fields in the jet bundle
$J^k \P$, and omit the indication of the bundle $\P$ (i.e. just
write $\Pr{k}$) when there is no risk of misunderstanding.

\subsection{Prolongation of general vector fields}
\label{sec:prolgenVF}

We will give some explicit formulas in the local coordinates
$(x,u,\a)$ introduced above. With the $(x,u)$ coordinates in $M$,
any vector field in $\wt{M}$ is written as $\wt{X} = \xi^i (\pa /
\pa x^i) + \vphi^a (\pa / \pa u^a) + B^m \ell_m$, with
$\{\xi^i,\vphi^a,B^m \}$ depending on $(x,u,g)$. Passing to the
restriction $\^X$ of $\wt{X}$ to $\^\G$, i.e. its expression in
$\^M$, and introducing also the local coordinates $\a$ in $\G$
(recall $\ell_m = \pa / \pa \a^m$), we have $\^X = \xi^i (x,u,\a )
(\pa / \pa x^i) + \vphi^a (x,u,\a) (\pa / \pa u^a) + \ B^m
(x,u,\a) (\pa / \pa \a^m)$.

As usual in considerations involving vector fields on jet bundles,
it will be convenient to work with evolutionary representatives
\cite{\symref}; we will consistently use these. The evolutionary
representative of $\^X$ is \beq\label{eq:Xv} X \ \equiv \ \^X_v \
= \ Q^a  \, {\pa \over \pa u^a} \ + \ P^m \, {\pa \over \pa \a^m}
\ , \eeq where $Q^a := \vphi^a - u^a_i \xi^i$, $P^m := B^m$.

The coordinate expression of the prolongation $X^{(k)} \in \X (J^k
\^M)$ of $X$ is given by the (standard) prolongation formula
\cite{\symref}. (As implied by the discussion in section
\ref{sec:totder} above, no prolongation of $\a^m$ components
appear). For the evolutionary representative $Y := (X^{(k)})_v =
X_v^{(k)}$ we get, with $Q^a_J = D_J Q^a$, \beq\label{eq:Y1} Y \ =
\ Q^a_J \, {\pa \over \pa u^a_J} \ + \ P^m \, {\pa \over \pa \a^m}
\ . \eeq

\subsection{Prolongation of gauged vector fields}

We are specially interested in a particular class of vector fields
in $\X_v (\wt{M})$, i.e. those for which \beq\label{eq:sepglob}
Q^a (x,u,g;u_x) \ = \ [\Psi (g)]^a_{\ b} \ \Q0^b (x,u;u_x) \ .
\eeq In the following we will refer to these as {\bf gauged vector
fields}.

Restricting gauged vector fields to $\^M$, and using local
coordinates $(x,u,\a)$, eq. (\ref{eq:sepglob}) becomes
\beq\label{eq:sep} Q^a (x,u,\a;u_x) \ = \ [K(\a)]^a_{\ b} \ \Q0^b
(x,u;u_x) \ ; \eeq here $K(\a)$ is the representation of the group
element $g (\a) = \exp (\a)$, i.e. $K(\a ) = \Psi [ \exp (\a)]$.
(We stress that (\ref{eq:sepglob}) and (\ref{eq:sep}) do not
constrain in any way the components $P^m$ of the vector fields
along the $\a^m$ variables; this will be of use below.)

Keeping in mind our discussion above about the total derivative
operators in $\^M^{(k)}$, see section \ref{sec:totder}, we obtain
immediately that \beq Q^a_J \ = \ D_J Q^a \ = \ [ K(\a )]^a_{\ b}
\ D_J \Q0^b \ . \eeq This implies that -- writing $\Q0_J = D_J
\Q0$ -- the prolongation $Y$ of the vector field $X$, see
\eqref{eq:Y1}, is given by \beq\label{eq:Y2}  Y \ = \ [K(\a)]^a_{\
b} \, \Q0^b_J \, {\pa \over \pa u^a_J} \ + \ P^m \, {\pa \over \pa
\a^m}  \ . \eeq

\medskip\noindent
{\bf Remark 3.} Let $X_0 = \rho_* X$ and $Y_0 = \rho^{(k)}_* Y$ be
the projection of the vector fields $X$ and $Y$ to the bundles,
respectively, $M$ and $J^k M$. Then we can state formally that
{\it $Y_0$ is the $\mu$-prolongation of $X_0$ for a suitable
$\mu$.} In fact, we have \beq X_0 = \rho_* X = [K(\a)]^a_{\ b} \,
\Q0^b \, {\pa \over \pa u^a} \ ; \ \ Y_0 = \rho^{(k)}_* Y =
[K(\a)]^a_{\ b} \, \Q0^b_J \, {\pa \over \pa u^a_J} \ . \eeq Thus
$X_0$ and $Y_0$ are the gauge transformed -- via the same gauge
transformation -- of vector fields $\=X_0$ and $\=Y_0$ such that
$\=Y_0$ is the ordinary prolongation of $\=X_0$; By Proposition 1
above, $Y_0$ is the $\mu$-prolongation of $X_0$ for a suitable
one-form $\mu$. Note this statement is only formal, as the $\a$
variables have no meaning when we work in $M$ and $J^k M$; in
order to make this into a real theorem, we will need to ``fix the
gauge'', as discussed below. Note also the relation between $X_0$
and $Y_0$ depends substantially on the assumption $X$ is a gauged
vector field. \EOR

\medskip\noindent
{\bf Remark 4.} The reason for the name ``gauged'' of vector
fields considered here is quite clear: the horizontal (for the
gauge fibration) component $Q^a \pa_a$ of these corresponds to
vector fields in $M$ on which we operate with an element of the
Lie group $G$, element which may vary for varying $x$ and
$u$.\footnote{Note that if we operate on $\eb$ by a $x$-dependent
change of frame, i.e. pass to a frame $\fb = (\fb_1 , ... , \fb_n
)$ with $\fb_a = T_a^{\ b} \eb_b$ where $T = (K^T)^{-1}$, then we
get $ \Phi = \vphi^a \eb_a = \vphi^a (T^{-1})_a^{\ b} \fb_b :=
\wt{\vphi}^a \fb_a$, i.e. the components of $\Phi$ in the new
frame are $\wt{\vphi}^a = K^a_{\ b} \vphi^b$. This point of view
is discussed elsewhere \cite{Gframe}.} \EOR

\section{Standard prolongation of vector fields in $\^M$ and
$\mu$-prolongation of vector fields in $M$}
\label{sec:muprolongation}

We showed above that the natural prolongation in gauge bundles is
naturally related (via the results in \cite{CGM}) to a
$\mu$-prolongation. However, $\lambda$- and $\mu$-prolongations
are usually \cite{Cicrho,CGM,GM,MuRom1,MuTMP,MuVigo,MRO,MuRom07}
defined with no use of auxiliary gauge variables; in our language
this will correspond to a gauge fixing.

In this section we will discuss how gauge fixing affects the
prolongation operation, and the relation between gauge-fixed
prolongation and $\mu$-prolongations.

\subsection{Sub-bundles defined by sections of the gauge bundle}

Earlier on we considered the augmented bundle $\^M$ and
correspondingly $J^k \^M$. Here we want to consider the subbundle
$\^M_\ga := \ga (M) \ss \^M$ defined by a section $\ga \ss
\Ga(\gaub )$ of the gauge bundle (hence by a section of the global
gauge bundle close to the reference section $\varpi$); we will
also consider $M_\ga^{(k)} := \ga (J^k M) \ss J^k \^M$.

In the $(x,u,\a)$ coordinates, $\^M_\ga$ is the set of points
$(x,u,\a)$ with $\a = A(x,u)$; similarly $\^M_\ga^{(k)}$ is the
set of points $(x,u,\a,u^{(1)},...,u^{(k)})$ again with $\a =
A(x,u)$. Note that $\^M_\ga \simeq M$, and correspondingly
$\^M_\ga^{(k)} \simeq J^k M$.

These submanifolds of the gauge bundle and of the jet gauge bundle
have a natural structure of fiber bundles (over $B$) themselves,
and can be seen as sub-bundles of $\^M$ and $J^k \^M$\footnote{We
stress this refers to the structure of bundles over $B$, i.e. --
referring to diagram \eqref{diag:star} -- to the $\^\pi_k$
projections; more care will be needed for what concerns the
structure of bundles over $\^M$, see section \ref{sec:prolgaufix}
below.}; we will thus also write $J^k \^M_\ga $ for
$\^M_\ga^{(k)}$. It will thus make sense to speak of vertical
vector fields in $\^M_\ga$ and $J^k \^M_\ga$ (referring implicitly
to these fiber bundle structures).

Given a section $\ga \in \Ga (\gaub )$, we will denote by
$\om^{(\ga)}$ the operator of restriction from $\^M$ to $\^M_\ga$,
and by $\rho^{(\ga)}$ the restriction of the projection $\rho :
\^M \to M$ to $\^M_\ga$. We also denote by $\om^{(\ga)}_k : J^k
\^M \to \^M_\ga^{(k)}$ and by $\rho^{(\ga)}_k : \^M_\ga^{(k)} \to
J^k M$ the lift of the maps $\om^{(\ga)}$ and $\rho^{(\ga)}$ to
maps between corresponding jet spaces of order $k$. Note that
while $\rho$ is of course not invertible, it follows from $M_\ga
\simeq M$ that $\rho^{(\ga)}$ is invertible, with
$(\rho^{(\ga)})^{-1} = \ga$. similarly, $\rho^{(\ga)}$ is
invertible.

We will summarize relations and maps between relevant fiber
bundles in the following diagram: \beq\label{diag:sectbundle}
 \matrix{ \^M & \mapright{\om^{(\ga)}} & \^M_\ga &
\mapright{\rho^{(\ga)}} & M \cr & & & & \cr \mapdown{j^{k} [\^M]}
& & \mapdown{j^{k} [\^M_\ga]} & & \mapdown{j^{k} [M]} \cr & & & &
\cr J^k \^M & \mapright{\om^{(\ga)}_k} & \^M_\ga^{(k)} &
\mapright{\rho^{(\ga)}_k} & J^k M \cr} \eeq

\medskip\noindent
{\bf Remark 5.} In physical terms, passing to consider $\^M_\ga$
rather than the full $\^M$, and $\^M_\ga^{(k)}$ rather than the
full $\^M^{(k)}$, corresponds to a {\bf gauge fixing}. \EOR

\subsection{Prolongations and gauge fixing}
\label{sec:prolgaufix}

When dealing with $M_\ga$, i.e. working in a fixed gauge, we
should think the gauge variables $\a$ as explicit functions of $x$
and $u$ (given by $A(x,u)$ identifying $\ga$); note this means
$D_i \a^m$ will now read as $D_i A^m (x,u)$ and thus will in
general give a nonzero function. This entails the jet structure of
$\^M_\ga^{(k)}$ is not the one inherited from the jet structure of
$\^M^{(k)}$ (which we denote by $j_k [\^M_\ga]$, the associated
prolongation operator being $\Pr{k} [\^M_\ga]$).

Let us now consider vector fields. Consider $X \in \X_v (\^M)$
written as in \eqref{eq:Xv}, and $X_\ga = \om^{(\ga)}_* X$ be its
restriction to $\^M_\ga \ss \^M$. Then $X_\ga$ is \beq\label{eq:Y}
X_\ga \ = \ Q^a_\ga \ (\pa / \pa u^a ) + P^m_\ga (\pa / \pa \a^m )
\, \eeq where the coefficients $Q_\ga , P_\ga$ are of course given
by \beq\label{eq:phitheta} Q^a_\ga  = [Q^a ]_{\a = A(x,u)} \ , \ \
P^m_\ga  = [P^m ]_{\a = A(x,u)} \ . \eeq

It should be noted that for arbitrary $X$ and $\ga$, the
submanifold $\ga$ is is general not invariant under $X_\ga$. More
precisely, with the notation (\ref{eq:phitheta}), we have:

\medskip\noindent
{\bf Lemma 1.} {\it Let $X$ be in the form \eqref{eq:Xv}. Then the
submanifold $\^M_\ga \ss \^M$ identified by $\a^m = A^m (x,u)$ is
invariant under $X_\ga$ if and only if \beq\label{eq:gainvar}
P^m_\ga \ = \ \( \pa A^m / \pa u^a \) \ Q^a_\ga  \ . \eeq}

\medskip\noindent
{\bf Proof.} By standard computation. Note that using the notation
in section \ref{sec:prolgenVF}, this also reads as $P^m = X(A)$ on
$\^M_\ga$. \EOP
\medskip

\medskip\noindent
{\bf Corollary 1.} {\it Given arbitrary smooth functions $Q^a
(x,u,\a)$, and an arbitrary section $\ga \in \Ga (\^M)$, there is
always a vector field $X_v \in \X_v (\^M )$ of the form $X_v = Q^a
\pa_a + P^m \pa_m$ and such that $X_v$ leaves $\^M_\ga$
invariant.}

\medskip\noindent
{\bf Proof.} In view of (\ref{eq:Y2}), the condition
(\ref{eq:gainvar}) reads also $P^m  = (\pa A^m / \pa u^a) Q^a$;
for any given $\ga$ the vector fields leaving $\^M_\ga$ invariant,
have arbitrary $Q^a$ and $P^m$ given by $P^m = (\pa A^m /\pa u^a)
Q^a  + \de P^m $ with $\de P^m$ arbitrary functions vanishing on
$\a = A(x,u)$. \EOP

\medskip\noindent
{\bf Remark 6.} The vector field $X_\ga$ projects in turn to a
vector field $W \in \X_v (M)$, $W = Q^a (x,u) (\pa / \pa u^a)$;
and conversely any such $W \in \X_v (M)$ lifts to a $W^\ga = X_\ga
\in \X_v (\^M_\ga)$, $ W^\ga = Q^a (x,u) [ (\pa / \pa u^a) + (
(\pa A^m / \pa u^a) (\pa / \pa \a^m)]$. \EOR
\medskip

The set of vector fields $X \in \X_v (\^M)$ which leave $\^M_\ga$
invariant, i.e. satisfy \eqref{eq:gainvar}, will be denoted as
$\X_v^{(\ga)} (\^M)$. If $X \in  \X_v^{(\ga)} (\^M)$, then $X_\ga$
is actually a vector field on $\^M_\ga$, and $X_\ga^{(k)}$ a
vector field on $J^k \^M_\ga$.

The diagram \eqref{diag:sectbundle} has a counterpart for these
vector fields. We will use for graphic convenience a simplified
notation with $ \^\X := \X_v^{(\ga)} (\^M)$, $\^\X_\ga := \X_v
(\^M_\ga)$, $\X := \X_v (M)$; $\^\X^{(k)} := \X_v^{(\ga)}
(\^M^{(k)})$, $\^\X_\ga^{(k)} := \X_v (\^M_\ga^{(k)})$, $\X^{(k)}
:= \X_v (M^{(k)})$. With this, \eqref{diag:sectbundle} yields:
\beq\label{diag:Z00} \matrix{ \^\X & \mapright{\om^{(\ga)}_*} &
\^\X_\ga & \mapright{\rho^{(\ga)}_*} & \X \cr & & & & \cr
\mapdown{\Pr{k} [\^M]} & & \mapdown{\Pr{k} [\^M_\ga]} & &
\mapdown{\Pr{k} [M]} \cr & & & & \cr \^\X^{(k)} &
\mapright{(\om^{(\ga)}_k)_*} & \^\X_\ga^{(k)} &
\mapright{(\rho^{(\ga)}_k)_*} & \X^{(k)} \cr} \eeq The operators
$\Pr{k} [\^M]$ and $\Pr{k} [\^M_\ga]$ should be understood with
the discussion of section \ref{sec:totder} in mind.

The diagram \eqref{diag:Z00} is in general {\it not } commutative.
We will now discuss how it can be made into a commutative one by
replacing $\Pr{k}[\^M_\ga]$ and $\Pr{k} [M]$ by, respectively,
suitable operators ${\tt \^P}_\ga^{(k)} : \X_v (\^M_\ga) \to \X_v
({\^M}^{(k)}_\ga)$ and ${\tt P}_\ga^{(k)} : \X_v (M) \to \X_v
(M^{(k)})$. That is, we want to identify ${\tt \^P}_\ga^{(k)}$ and
${\tt P}_\ga^{(k)}$ yielding, for a given $X \in \X_v^{(\ga)}
(\^M)$,  \beq\label{diag:ZZ}  \matrix{ X &
\mapright{\om^{(\ga)}_*} & X_\ga & \mapright{\rho^{(\ga)}_*} & W
\cr & & & & \cr \mapdown{\Pr{k} [\^M]} & & \mapdown{{\tt
\^P}_\ga^{(k)}} & & \mapdown{{\tt P}_\ga^{(k)}} \cr & & & & \cr
X^{(k)} & \mapright{(\om^{(\ga)}_k)_*} & X_\ga^{(k)} &
\mapright{(\rho^{(\ga)}_k)_*} & Y \cr} \eeq

\subsection{Twisted differential operators in $\^M_\ga$}
\label{sec:twistdiffop}

Let us first discuss the left-hand side of the diagram
\eqref{diag:ZZ}. We will write $K_\ga$ for $K(\a)$ computed on $\a
= A(x,u)$.

In general we obtain different results by changing the order in
which the prolongation and the gauge fixing operations are
performed.

\medskip\noindent
{\bf Remark 7.} This is immediately seen by considering a vector
field in the form \eqref{eq:Xv}, \eqref{eq:sep}. We have of course
$X_\ga = Q_\ga^a (\pa / \pa u^a) + P^m_\ga (\pa / \pa \a^m)$; as
$Q_\ga = K_\ga \Q0$, the prolongation of $X_\ga$ is $(X_\ga)^{(k)}
= \=\psi^a_J (\pa / \pa u^a_J) + P^m_\ga (\pa / \pa \a^m)$, with $
\=\psi^a_J = (D_J Q_\ga^a) = D_J [(K_\ga)^a_{\ b} \Q0^b]$. On the
other hand, the prolongation of $X$ is $X^{(k)} = (D_J Q^a)(\pa /
\pa u^a_J) + P^m_\ga (\pa / \pa \a^m)$ with $D_J Q^a = D_J (K^a_{\
b} \Q0^b) = K^a_{\ b} D_J (\Q0^b);$ hence gauge fixing after
prolongation yields $(X^{(k)})_\ga = \psi^a_J (\pa / \pa u^a_J) +
P^m_\ga (\pa / \pa \a^m)$ with $ \psi^a_J = (K_\ga)^a_{\ b}  D_J
(\Q0^b)$. Needless to say, in general $\psi^a_J \not= \=\psi^a_J$.
\EOR
\medskip

Denote by $\de_\ga$ the operator fixing the gauge to $\ga$. We
will look for ``twisted differential operators''
$\nabla_i^{(\ga)}$ (we will also write $\nabla_i$ for short) such
that \beq\label{eq:nabla} \de_\ga \ [ D_J (Q^a )] \ = \
\nabla_J^{(\ga)}
\[ \de_\ga (Q^a ) \] \eeq when $Q$ is of the form \eqref{eq:sep}.
The operator ${\tt \^P}_\ga$ will then be the ``twisted
prolongation'' obtained by replacing $D_J$ with
$\nabla_J^{(\ga)}$.

Let us define matrices $R^{(\ga)}_i$ by \beq\label{eq:fpRsect}
R^{(\ga)}_i (x,u,u_x) \ = \ (D_i K_\ga ) \ K_\ga^{-1} \ .  \eeq

\medskip\noindent
{\bf Lemma 2.} {\it The matrices $R_i^{(\ga)} (x,u,u_x)$ satisfy
the horizontal Maurer-Cartan equation \beq\label{eq:MCRsect} D_i
\, R^{(\ga)}_j \ - \ D_j \, R^{(\ga)}_i \ + \ [ R^{(\ga)}_i ,
R^{(\ga)}_j ] \ = \ 0 \ . \eeq }

\medskip\noindent
{\bf Proof.} It follows from (\ref{eq:fpRsect}) that $ D_i
R^{(\ga)}_j  = (D_i D_j S) S^{-1} - R^{(\ga)}_i R^{(\ga)}_j$.
Recalling $[D_i,D_j]=0$, we get $ D_i R^{(\ga)}_j - D_j
R^{(\ga)}_i = - [R^{(\ga)}_i , R^{(\ga)}_j ]$. \EOP
\medskip

We will define the operators $\nabla_i^{(\ga)}$ as
\beq\label{eq:nabladef} \nabla_i^{(\ga)} \ := \ D_i \ - \
R_i^{(\ga)} \ . \eeq It follows from Lemma 2 that $[\nabla_i ,
\nabla_j] = 0$. For a multiindex $J = (j_1,...,j_m)$, the
operators $\nabla_J^{(\ga)}$ are defined as $\nabla_J =
\nabla_1^{j_1} ... \nabla_m^{j_m}$; this is well defined in view
of $[\nabla_i , \nabla_j] = 0$.

\medskip\noindent
{\bf Lemma 3.} {\it Let $Q^a = K^a_{\ b} \Q0^b$. The operators
$\nabla_i^{(\ga)}$ defined in \eqref{eq:nabladef} satisfy
\beq\label{eq:nablai} \de_\ga \, [ D_i (Q^a) ] \ = \
\nabla_i^{(\ga)} \, [Q^a_\ga] \eeq for all $i$, and hence
\eqref{eq:nabla} for all multi-indices $J$.}

\medskip\noindent
{\bf Proof.} We can proceed by direct computation; we will omit
indices in intermediate formulas for ease of notation. For $Q = K
\Q0$ we have immediately $D_i Q = K [D_i ( \Q0)]$ and hence
\beq\label{eq:prN1} \de_\ga \, [ D_i (Q^a) ] \ = \ (K_\ga)^a_{\ b}
\ D_J \Q0^b \ . \eeq On the other hand, $\nabla_i (Q_\ga)  = D_i
[(K_\ga) \Q0 ] - [R_i^{(\ga)}] (K_\ga) \Q0 = K_\ga D_i \Q0 + [D_i
K_\ga] \Q0 - [R_i^{(\ga)}] (K_\ga) \Q0$. Recalling the definition
of $R_i^{(\ga)}$, we have $$ [R_i^{(\ga)}] (K_\ga) \Q0 = (D_i
K_\ga ) K_\ga^{-1} K_\ga \Q0 = (D_i K_\ga ) \Q0 \ , $$ and hence
\beq\label{eq:prN2} \nabla_i \, [Q_\ga^a]  \ = \ (K_\ga)^a_{\ b} \
[D_i (\Q0^b)] \ ; \eeq comparing this and \eqref{eq:prN1} we
obtain \eqref{eq:nablai}.

In order to see this implies \eqref{eq:nabla} it suffices to note
we have actually proved \beq \nabla_i \ [K_\ga \, \Q0 ] \ = \
K_\ga \ [D_i \Q0 ] \ ; \eeq applying this repeatedly we obtain
\eqref{eq:nabla}. Alternatively, one can explicitly compute
$(\nabla_i \nabla_j Q_\ga) $ and check this is equal to $\de_\ga (
D_i D_j Q )$. \EOP
\medskip

As anticipated, the operator ${\tt \^P}_\ga$ will be the ``twisted
prolongation'' obtained by replacing $D_J$ with $\nabla_J$.

\medskip\noindent
{\bf Lemma 4.} {\it Let ${\tt \^P}_\ga $ be the (twisted
prolongation) operator associating to any vector field $X_\ga =
Q^a_\ga (\pa / \pa u^a) + P^m_\ga (\pa / \pa \a^m)$ in $\X_v
(\^M_\ga)$ the vector field $Y_\ga = \eta^a_J (\pa / \pa u^a_J) +
P^m (\pa / \pa \a^m)$ in $\X_v (\^M_\ga^{(k)})$ with coefficients
$\eta^a_J := \nabla_J^{(\ga)} Q^a_\ga$. Then the
left-hand side of diagram \eqref{diag:ZZ} is commutative.}

\medskip\noindent
{\bf Proof.} See the explicit computations in the proof to Lemma
3. \EOP

\medskip\noindent
{\bf Lemma 5.} {\it Let ${\tt P}_\ga $ be the (twisted
prolongation) operator associating to any vector field $W =
Q^a_\ga (\pa / \pa u^a)$ in $\X_v (M)$ the vector field $Y =
\eta^a_J (\pa / \pa u^a_J)$ in $\X_v (M^{(k)})$ with coefficients
$\eta^a_J := \nabla_J^{(\ga)} Q^a_\ga$. Then the right-hand side
of diagram \eqref{diag:ZZ} is commutative.}

\medskip\noindent
{\bf Proof.} This ${\tt P}_\ga $ is nothing else than the
restriction of ${\tt \^P}_\ga $ to the components along $M^{(k)}$.
It is obvious (by construction) that ${\tt P}_\ga \circ
\rho_*^{(\ga)} = (\rho_k^{(\ga)})_* \circ {\tt \^P}_\ga$. \EOP

\subsection{The main results}
\label{sect:main}

We are now ready to state and prove our main results, which will
actually just collect results appearing in the previous Lemmas;
these will make the formal statement in Remark 3 into precise
ones.

We will introduce, in order to state our result in a compact form,
operators $\tau^{(\ga)} := \rho^{(\ga)} \circ \om^{(\ga)}$,
$\tau^{(\ga)} : \^M \to M$; and correspondingly $\tau^{(\ga)}_k :=
\rho^{(\ga)}_k \circ \om^{(\ga)}_k$, $\tau^{(\ga)}_k : \^M^{(k)}
\to M^{(k)}$.

\medskip\noindent
{\bf Theorem 1.} {\it The twisted prolongation operator ${\tt
P}_\ga^{(k)}$ is uniquely defined by the requirement that
$(\tau^{(\ga)}_k)_* \circ (\Pr{k} [\^M]) = {\tt P}_\ga^{(k)} \circ
\tau^{(\ga)}_*$.

Moreover, ${\tt P}_\ga^{(k)}$ corresponds to the
$\mu$-prolongation operator of order $k$ with $\mu = [\d
\Psi(\ga)] \Psi(\ga^{-1})$. With the local coordinates $(x,u,\a)$,
this corresponds to $\mu = \La_i \d x^i$ where $\La_i = -
R_i^{(\ga)} = - (D_i K_\ga) K_\ga^{-1} = K_\ga (D_i K_\ga^{-1}
)$.}

\medskip\noindent
{\bf Proof.} The first part of the statement just summarizes the
discussion in section \ref{sec:twistdiffop}. As for the second
part, it follows at once considering the definitions of ${\tt
P}_\ga^{(k)}$, of $\nabla_i^{(\ga)}$ and of $R_i^{(\ga)}$, and
comparing with the $\mu$-prolongation formula \eqref{eq:muprol}.
\EOP

\medskip\noindent
{\bf Theorem 2.} {\it Let $Y \in \X_v (J^k M)$ be the $k$-th
$\mu$-prolongation of $W \in \X_v (M)$, with $\mu \in \La^1 (J^1 M
, \psi (\G))$ given in coordinates by $\mu = \La_i (x,u,u_x) \d
x^i$; let $\gaub$ be the local gauge bundle over $M$ with fiber
$\G$. Then:
\par\noindent {\tt (i)} there is a section $\ga \in \Ga (\gaub )$
such that $ Y = {\tt P}_\ga (W)$.
\par\noindent {\tt (ii)} there is $X \in \X_v
(\^M)$ such that (\ref{diag:ZZ}) applies.
\par\noindent {\tt (iii)} The matrix function $K_\ga (x,u,u_x)$
satisfies $D_i K_\ga = - \La_i K_\ga$.}

\medskip\noindent
{\bf Proof.} The theorem states that any $\mu$-prolongation can be
obtained locally through the construction considered here. Part
{\tt (i)} is obvious given the identification of
$\mu$-prolongation and the ${\tt P}_\ga$ operators, see Lemma 5.
Part {\tt (ii)} follows at once from Lemma 4 and Lemma 5. Part
{\tt (iii)} follows from the relation $\La_i = - R_i^{(\ga)} = -
(D_i K_\ga) K_\ga^{-1}$ already considered above (note $K_\ga$ is
surely invertible as it is the representation of an element of the
Lie group $G$). \EOP
\bigskip

It is maybe worth commenting on the geometrical meaning of our
main results.

We have shown that $\mu$-symmetries in the phase bundle can be
understood as ordinary symmetries in the augmented phase bundle,
restricted to a section of the gauge bundle and mapped to the
standard phase bundle.

In other words, we have obtained that the $\mu$-prolongation
operator appears if we are insisting in restricting our analysis
to the phase bundle $M$ (or to the subbundle $\^M_\ga \ss \^M$
seen as an image of $M$ under the gauge map $\ga$ embedding it
into $\^M$) rather than to the full gauge bundle $\^M$.

The fact we are considering projections of vector fields in
$\^M_\ga \ss \^M$ and ${\^M}_\ga^{(k)} \ss \^M^{(k)}$ to vector
fields in $M$ and $M^{(k)}$ makes that the relation between basic
vector fields and prolonged ones is not the natural one, described
by the prolongation operator, but is the ``twisted'' one described
by the $\mu$-prolongation operator. See also the discussion in the
Appendix.

We would also like to stress that gauge fields and hence variables
can also be seen -- like in standard gauge theories -- as indexing
reference frames. The restriction to a section of the gauge bundle
corresponds thus to fixing a gauge and hence a reference frame
(not equivalent to the original, ``natural'', one if the section
is nontrivial); projection to the phase bundle corresponds to
loosing track of the change of reference frame. See also
\cite{Gframe}.

Finally we mention that our construction in the augmented phase
bundle corresponds, when working in the phase bundle alone, to the
fact that (locally, and globally when we deal with topologically
trivial phase bundles) $\mu$-symmetries can be transformed into
standard ones by a gauge transformation \cite{CGM}.

\section{Examples I. Abelian groups}

In this section we will provide some very simple examples
illustrating our results; we will consider $B = \R$ and $U = \R^2$
(see e.g. \cite{CGM} for the reasons making the case $U = R^1$
less interesting in this context); we will consider one-parameter
(hence abelian; see next section for non-abelian examples) Lie
groups $G$, its Lie algebra being $\G$ with generator $\ell$. The
gauge variable will be denoted as $\a$. We will use the notation
$(u,v)$ for the coordinates $(u^1,u^2)$ in $U$; we denote by $L$
the representation $\psi (\ell)$ of the generator $\ell$.

We will consider second order prolongations (i.e. $k=2$), and
write the vector field $Y$, see \eqref{diag:ZZ}, in the form
\beq\label{eq:exaY} Y = \eta^a (\pa / \pa u^a) + \eta^a_x (\pa /
\pa u^a_x) + \eta^a_{xx} (\pa / \pa u^a_{xx}) \ ; \eeq the
required relation between its coefficients will then be
\beq\label{eq:exarecrel} \eta_x = D_x \eta + \La \eta \ , \ \
\eta_{xx} = D_x \eta_x + \La \eta_x \ . \eeq

Examples 1 through 4 concern Theorem 1, while Examples 5 and 6
deal with Theorem 2.

\subsection{Example 1.}

Let us first consider the case where $G = R$ acts in $U$ via
$$ T(g) \ = \ \pmatrix{1&g\cr0&1\cr} . \ $$ In this case we
have immediately $$ L = \pmatrix{0&1\cr0&0\cr}  , \ \ K  \ = \
T[e^{\a L}] \ = \ \pmatrix{1&\a \cr0&1\cr} \ ; \
K^{-1} \ = \ \pmatrix{1&- \a \cr0&1\cr} \ . $$

Let us consider the gauged vector field $X$ of the form
\eqref{eq:sep} with $K(\a)$ as above and $\Q0 = (2 u , v)$; the
gauge section $\ga$ to be considered will be the one identified by
$$ \a \ = \ A(x,u,v) \ := \ u \ . $$ The invariance of $\^M_\ga$
requires then, see \eqref{eq:gainvar} above, $ P = [X(A)] = (2 u +
\a v)$. We will thus be considering the vector field
$$ X  \ = \ (2 u + \a v) (\pa/\pa u) \ + \ v \,
(\pa/\pa v) \ + \ (2 u + \a v) \, (\pa/\pa \a) \ . $$
The restriction of $X$ to $\^M_\ga$ is
$$ X_\ga  \ = \ (2 u + u v) (\pa/\pa u) \ + \ v \,
(\pa/\pa v) \ + \ (2 u + u v) \, (\pa/\pa \a) \ , $$ and the
projection of $X_\ga$ to $M$ is simply
$$ W \ = \ (2 u + u v) (\pa/\pa u) \ + \ v \,
(\pa/\pa v) \ . $$

As for second prolongations, it follows from general formulas that
$$ \begin{array}{rl}
X^{(2)} \ =& X \ + \ (2 u_x + \a v_x) (\pa/\pa u_x) \ + \ v_x \,
(\pa/\pa v_x) \\ &  + \ (2 u_{xx} + \a v_{xx}) (\pa/\pa u_{xx}) \ + \
v_{xx} \, (\pa/\pa v_{xx}) \ . \end{array} $$ Restricting this to
$\^M_\ga^{(2)}$ and projecting to $M^{(2)}$ yields
$$ \begin{array}{rl} Y \ =& W \ + \ (2 u_x + u v_x) (\pa/\pa u_x) \ + \ v_x \,
(\pa/\pa v_x) \\ &  + \ (2 u_{xx} + u v_{xx}) (\pa/\pa u_{xx}) \ +
\ v_{xx} \, (\pa/\pa v_{xx}) \ . \end{array} $$ Writing this in
the form \eqref{eq:exaY} we have
$$ \eta = \pmatrix{2 u + u v \cr v \cr} \ , \ \
\eta_x = \pmatrix{2 u_x + u v_x \cr
v_x \cr}\ , \ \ \eta_{xx} = \pmatrix{2 u_{xx} + u v_{xx} \cr
v_{xx}\cr} \ . $$

According to our general theorem, these should satisfy the
recurrence formula \eqref{eq:exarecrel} with $\La = - R_x^{(\ga)}
= - (D_x K_\ga) K_\ga^{-1}$. With our choices we have
$$ \La = - R_x^{(\ga)} = - \pmatrix{0 & u_x \cr 0 & 0 \cr} \pmatrix{1
& - u \cr 0 & 1 \cr} =  \pmatrix{0 & - u_x \cr 0 & 0 \cr} \ . $$
It is immediate to check the $\eta^a_J$ satisfy
\eqref{eq:exarecrel} with this $\Lambda$.

\subsection{Example 2.}

Let us now the case where $G = R$ acts in $U$ as above, and $\ga$
is also as above (so that the $K_\ga$, $R^{(\ga)}$ and $\La$ are
the same as before), but $\Q0 = (- v , u)$. Now the condition $P =
X(A)$ yields $P = (- v + \a u)$ That is, we have
$$ X  \ = \ (- v + \a u) (\pa/\pa u) \ + \ u \,
(\pa/\pa v) \ + \ (\a u - v ) \, (\pa/\pa \a) \ . $$
The restriction of $X$ to $\^M_\ga$ and its projection to $M$ are
$$ X_\ga  \ = \ (u^2 - v) (\pa/\pa u) \ + \ u \,
(\pa/\pa v) \ + \ (u^2 - v ) \, (\pa/\pa \a) \ ; $$
$$ W  \ = \ (u^2 - v) (\pa/\pa u) \ + \ u \,
(\pa/\pa v) \ . $$

Let us consider again second prolongations. Now
$$ \begin{array}{rl}
X^{(2)} \ =&  X \ + \ (- v_x + \a u_x) (\pa/\pa u_x) \ + \ u_x \,
(\pa/\pa v_x) \\ & + \ (- v_{xx} + \a u_{xx}) (\pa/\pa u_{xx}) \ +
\ u_{xx} \, (\pa/\pa v_{xx})  \ . \end{array} $$ Restricting this
to $\^M_\ga^{(2)}$ and projecting to $M^{(2)}$ yields $Y$; with
the notation \eqref{eq:exaY} this corresponds to
$$ \eta = \pmatrix{u^2 - v \cr u \cr} \ , \ \
\eta_x = \pmatrix{u u_x - v_x \cr u_x \cr}\ , \ \ \eta_{xx} =
\pmatrix{u u_{xx} - v_{xx} \cr u_{xx}\cr} \ . $$ The matrix $\La$
is as above, and one checks easily the required recursion
relations \eqref{eq:exarecrel} are satisfied.

\subsection{Example 3.}

Let us now consider $G = SO(2) \simeq S^1$ acting in $U = R^2$
through its standard representation, $$ T(g) \ = \ \pmatrix{\cos
(g) & - \sin (g) \cr \sin (g) & \cos (g) \cr} \ . $$ In this case
$$ L \ = \ \pmatrix{0 & - 1 \cr 1 & 0 \cr} \ ; \ \
K (\a) = e^{\a L} = \pmatrix{\cos (\a) & - \sin (\a) \cr \sin (\a)
& \cos (\a) \cr} \ . $$ We will once again consider $\ga$
identified by $\a = u$; it follows by the usual formulas that
$$ \La = - R_x^{(\ga)} = - (D_x K_\ga) K_\ga^{-1} \ = \
\pmatrix{0 & u_x \cr - u_x & 0 \cr} \ . $$

We will consider the gauged vector field with $Q = K \Q0$ and the
same $\Q0$ considered in Example 1. This yields $Q = (2 u \cos \a
- v \sin \a, v \cos \a +2 u \sin \a )$, and $P = X(A)$ yields $P =
2 u \cos \a - v \sin \a$. With the by now usual computations, we
get $ W = [2 u \cos (u) - v \sin (u)] (\pa / \pa u) + [v \cos (u)
+ 2 u \sin (u)] (\pa / \pa v)$; and working in the usual way with
second prolongations we get
$$ \begin{array}{l}
\eta = \pmatrix{2 u \cos (u) - v \cos (u)  \cr v \cos (u) + 2 u
\cos (u)  \cr} \ , \ \ \eta_x = \pmatrix{2 u_x \cos (u) - v_x \cos
(u)  \cr v_x \cos (u) + 2 u_x \cos (u)  \cr} \ , \\ \eta_{xx} =
\pmatrix{2 u_{xx} \cos (u) - v_{xx} \cos (u)  \cr v_{xx} \cos (u)
+ 2 u_{xx} \cos (u)  \cr} \ . \end{array} $$ It is again immediate
to check that these satisfy the required recursion relation with
the $\La$ computed above.

\subsection{Example 4.}

We consider the same $G$-action and gauge section as above, so
that the $K$, $R_x$ and $\La$ are the same as in Example 3; but
choose now $Q = K \Q0$ with the $\Q0$ considered in Example 2,
$\Q0 = (- v , u)$. In this case
$$ \begin{array}{l}
\eta = \pmatrix{-v \cos (u) -u \cos (u)  \cr u \cos (u) - v \cos (u) \cr}
\ , \ \ \eta_x = \pmatrix{- v_x \cos (u) - u_x \cos (u)  \cr u_x \cos (u)
- v_x \cos (u)  \cr} \ , \\
\eta_{xx} = \pmatrix{-v_{xx} \cos (u) - u_{xx} \cos (u)  \cr u_{xx} \cos (u)
- v_{xx} \cos (u)  \cr} \ . \end{array} $$
It is again immediate to check that \eqref{eq:exarecrel} are satisfied.

\subsection{Example 5.}

Consider the vector field $Y$ in the form \eqref{eq:exaY} with
$$ \eta = \pmatrix{0 \cr v \cr} \ , \ \ \eta_x = \pmatrix{u_x v
\cr v_x \cr} \ , \ \ \eta_{xx} = \pmatrix{u_{xx} v + 2 u_x v_x \cr
v_{xx} \cr} \ . $$ It is immediate to see these satisfy
\eqref{eq:exarecrel} with
$$ \La \ = \ \pmatrix{0 & u_x \cr 0 & 0 \cr} \ ; $$
we already know from Examples 1 and 2 the form of the $K$ giving
such a $\La$, but let us pretend we do not. We have then to solve
$\La = - (D_x K_\ga) K_\ga^{-1}$ as an equation for $K_\ga$;
recall we look for solutions such that $K_\ga$ belongs to a
one-parameter (hence abelian) Lie group. Note that we can write
$\La = D_x P$ for a matrix $P$. We can rewrite the equation
linking $\La$ and $K$ as $ D_x P \ = \ - \, D_x (\log K_\ga)$,
with solution $K_\ga = \exp(- P)$ (the arbitrary constant matrix
for $K_\ga$ can be embodied in the arbitrary constant matrix for
$P$). In our case we get easily $K_\ga$; it is immediate to see
the resulting $K_\ga$ as the restriction to $\a = u$ of a matrix
$K (\a)$,
$$ K_\ga = \pmatrix{1&-u\cr0&1\cr} \ ; \ \
K (\a ) = \pmatrix{1&- \a \cr0&1\cr} \ . $$ Note these allow to
build the diagram \eqref{diag:ZZ} by working backwards; that is,
we act on the components $\eta^a_J$ of $Y$ by the matrix $K
K_\ga^{-1}$, thus obtaining $X^{(2)} = \psi^a_J (\pa / \pa u^a_J)
+ P (\pa / \pa \a)$ with $P = X(u) = [(u-\a) v]$ and
$$ \begin{array}{l}
\psi = \pmatrix{(u - \a) v \cr v \cr} \ , \ \
\psi_x = \pmatrix{u_x v + (u-\a) v_x \cr v_x \cr} \ , \\
\psi_{xx} = \pmatrix{u_{xx} v + 2 u_x v_x + (u-\a) v_{xx} \cr
v_{xx} \cr } \ . \end{array} $$ It is immediate to check this is
the standard second prolongation in $\^M^{(2)}$ of $X = [(u - \a)
v] (\pa / \pa u) + v (\pa / \pa v) + [(u - \a) v] (\pa / \pa \a)$.

\subsection{Example 6.}

Consider the vector field $Y$ in the form \eqref{eq:exaY} with
$$ \eta = \pmatrix{1 \cr 0 \cr} \ , \ \ \eta_x = \pmatrix{0
\cr v_x \cr} \ , \ \ \eta_{xx} = \pmatrix{0 \cr v_{xx} \cr} \ . $$
It is immediate to see these satisfy \eqref{eq:exarecrel} with
$$ \La \ = \ \pmatrix{0 & 0 \cr v_x & 0 \cr} \ . $$
Proceeding as above, and choosing as $\ga$ the section $\a = - v$
of the gauge bundle, we get
$$ K_\ga \ = \ \pmatrix{1&0\cr - v & 1\cr} \ , \ \ K(\a) \ = \
\pmatrix{1&0\cr \a & 0 \cr} \ . $$

We thus obtain $X^{(2)} = \psi^a_J (\pa / \pa u^a_J) + P (\pa /
\pa \a)$ with $P = -(\a+v)$ and
$$ \psi = \pmatrix{1 \cr \a + v \cr} , \
\psi_x = \pmatrix{0 \cr v_x \cr} , \ \psi_{xx} = \pmatrix{0 \cr
v_{xx} \cr } \ ; $$ i.e. the second prolongation of $X =  (\pa /
\pa u) + (\a + v) (\pa / \pa v) - (\a + v) (\pa / \pa \a)$.

\section{Examples II. Non-abelian groups: SU(2)}

The examples considered so far all concern very simple actions of
a one-generator, hence abelian, Lie group. It would not be
difficult to extend these to actions of $k$-generators abelian
groups; but for physical applications one should rather consider
non-abelian Lie groups such as e.g. the rotation or unitary
groups.

In this section we will consider the group $G=SU(2)$
acting\footnote{We have so far used real vector spaces; in order
to avoid converting at this point to complex ones, we will use a
real representation of $G=SU(2)$, acting in $R^4$ rather than in
$C^2$.} in $R^4 \simeq C^2$; we will consider very simple vector
fields and section, but still will have to set down rather complex
formulas. We apologize to the reader for such unavoidable
complexities; computations were performed in {\tt Mathematica},
which also shows that our formalism can be readily implemented
with a symbolic manipulation program.

Examples 7 and 8 concern Theorem 1, while Examples 9 and 10 deal
with Theorem 2.

\subsection{SU(2) algebra and group action; lambda matrices}
\label{sect:su2algebra}

We will consider generators $$ \begin{array}{l} L_1 = T(\ell_1) =
\pmatrix{0&1&0&0\cr -1&0&0&0\cr 0&0&0&1\cr 0&0&-1&0\cr} , L_2 =
T(\ell_2) = \pmatrix{0&0&0&1\cr 0&0&1&0\cr 0&-1&0&0\cr
-1&0&0&0\cr} , \\ { } \\  L_3 = T(\ell_3) = \pmatrix{0&0&1&0\cr
0&0&0&-1\cr -1&0&0&0\cr 0&1&0&0\cr} \ . \end{array} $$ These
satisfy the $su(2)$ relations $ L_i L_j = \eps_{ijk} L_k -
\de_{ij} I$; here $I$ is the four dimensional identity matrix, and
summation over repeated indices is implicit here and in the
following. These relations also imply $[L_i , L_j ] = 2 \eps_{ijk}
L_k$, $\{ L_i , L_j \} = - 2 \de_{ij} I$.

In the following we will need to compute the group element $g$
corresponding to $g = \exp(\ell)$ for $\ell$ an element of the
algebra. Consider a generic element of the algebra $\ell = \a^k
\ell_k$, and correspondingly a generic matrix
$$ L \ = \ \a^1 L_1 + \a^2 L_2 + \a^3 L_3 $$
(in the following we write all indices as lower ones in order to
avoid confusion with exponents in the computations). This is
written explicitly as
$$ L \ = \ \pmatrix{0&\a_1&\a_3&\a_2\cr -\a_1&0&\a_2&-\a_3\cr
-\a_3&-\a_2&0&\a_1\cr -\a_2&\a_3&-\a_1&0\cr} $$ and its square is
given by $ L^2 = - |\a|^2 I$, with $|\a| := \sqrt{\a_1^2 + \a_2^2
+ \a_3^2}$. Thus higher powers of $L$ satisfy \beq\label{eqsu2:L}
L^{2k} \ = \ (-1)^k \ |\a|^{2k} \ I \ ; \ \ L^{2k+1} \ = \ (-1)^k
\ |\a|^{2k} \ L \ . \eeq We have to compute $ e^L :=
\sum_{k=0}^\infty \ \fracor{L^k}{k!}$; it follows from
\eqref{eqsu2:L} that
$$ e^L \ = \ \sum_{k=0}^\infty \ \left[ \frac{(-1)^k |\a|^{2k}}{(2k)!} \, I \ + \
\frac{(-1)^k |\a|^{2k}}{(2k+1)!} \, L \right] \ ; $$ recognizing
the Taylor expansions of trigonometric functions, this reads
$$ e^L \ = \ \cos (|\a| ) \, I \ + \ |\a|^{-1} \,
\sin (|\a| ) \, L \ . $$ Recalling the definitions of $|\a|$ and
$L$, we have
$$ \begin{array}{rl} K(\a) \ :=& \ T [\exp (\a_1 L_1 + \a_2 L_2 +
\a_3 L_3)] \\ & = \ \cos (|\a|) \, I \ + \ \sin (|\a|) \,
[(\a_1/|\a|) L_1 + (\a_2 / |\a|) L_2 + (\a_3 / |\a|) L_3] \ .
\end{array} $$ We will introduce the matrix $ J :=
[(\a_1/|\a|) L_1 + (\a_2 / |\a|) L_2 + (\a_3 / |\a|) L_3] =
|\a|^{-1} L $; note $J^2 = - I$. With this, \beq\label{eqsu2:K}
\begin{array}{l} K(\a) \ = \ \cos (|\a|) \, I \ + \ \sin (|\a|) \, J \ ; \\
K^{-1} (\a) \ = \ \cos (|\a|) \, I \, - \, \sin (|\a|) \, J \ .
\end{array} \eeq

These formulas allow to give the general form of gauged vector
fields for this $G$ action: under the action of $K(\a)$, a vector
field of components $\Theta^a$ is transformed into a vector field
of components $\Phi^a = K^a_{\ b} \Theta^b$ given by
\beq\label{eqsu2:gauged} \Phi \ = \ \cos (|\a|) \ \Theta \ + \
\frac{\sin(|\a|)}{|\a|} \ \pmatrix{
   \a_1 \th_2 + \a_3 \th_3 + \a_2 \th_4 \cr
 - \a_1 \th_1 + \a_2 \th_3 - \a_3 \th_4 \cr
 - \a_3 \th_1 - \a_2 \th_2 + \a_1 \th_4 \cr
 - \a_2 \th_1 + \a_3 \th_2 - \a_1 \th_3 \cr} \ . \eeq
 This also fully describes the set of vector fields in $\R^4$
 which are gauged vector fields under the action of the presently
 considered representation of $SU(2)$.

Our formalism also requires to consider matrices $\La = - (D_x
K_\ga) K_\ga^{-1}$. It is possible, with standard but rather
tedious computations, to obtain the explicit form of these
starting from \eqref{eqsu2:K}, and restricting to the section
$\ga$ identified by $\a_m = A_m (x,u,v,w,z)$; using the notation
$\om = |\a|_\ga = \sqrt{A_1^2 + A_2^2 + A_3^2}$, the final result
turns out to be
$$ \La \ = \ \cos^2 (\om) \, M_1 \ + \ \sin^2 (\om) \, M_2 \ + \
\sin (\om) \, \cos (\om) \, M_3 \ , $$ where the $M_i$ are
four-dimensional matrices. These can be written in terms of the
matrix
$$ \L \ = \ \pmatrix{0& -A_1& -A_3& -A_2\cr A_1& 0& -A_2& A_3\cr A_3& A_2& 0&
-A_1\cr A_2& -A_3& A_1& 0\cr} \ , $$ of the matrix $D_x \L$ with
entries $(D_x \L)_{ij} = D_x [(\L)_{ij}]$, and of the
skew-symmetric matrix $\M$ with entries
$$ \begin{array}{ll}
 \M_{12} = A_3 D_x A_2 - A_2 D_x A_3 , &
 \M_{13} = A_2 D_x A_1 - A_1 D_x A_2 , \\
 \M_{14} = A_1 D_x A_3 - A_3 D_x A_1 , &
 \M_{23} = A_1 D_x A_3 - A_3 D_x A_1 , \\
 \M_{24} = A_1 D_x A_2 - A_2 D_x A_1 , &
 \M_{34} = A_3 D_x A_2 - A_2 D_x A_3 , \end{array}$$
in the form
$$ \begin{array}{l}
M_1 \ = \ \om^{-1} \, (D_x \om) \ \L \ ; \\
M_2 \ = \ \om^{-2} \ \M \ + \ \om^{-1} \, (D_x \om) \ \L \ ; \\
M_3 \ = \ - \, \om^{-2} \, (D_x \om) \ \L \ + \ \om^{-1} \ (D_x
\L) \ . \end{array} $$

\subsection{Example 7.}

We can now consider a concrete example, i.e. a specific vector
field to be $\mu$-prolonged and a specific section $\gamma$.

We will use coordinates $(u,v,w,z)$ for the space $U = \R^4$ of
dependent variables, and restrict to the subset $|u| < 1$. We
choose a vector field $X_0$ depending on $\a$ and acting in $U$.
This will be $$ X_0 \ = \ - u \cos(|\a|) \pa_u \ + \
\(\sin(|\a|)/|\a|\) \ \( \a^1 u \pa_v \, + \, \a^3 u \pa_w \, + \,
\a^2 u \pa_z \) \ , $$ which is obtained from \eqref{eqsu2:gauged}
for $\Theta = (-u,0,0,0)$.

As for the section $\ga$ we choose the one identified by
\beq\label{eqsu2:A} A^1 = B u \ , \ A^2 = 0 \ , \ A^3 = B
\sqrt{1-u^2} \ , \eeq with $B \not= 0$ an arbitrary real constant;
in the following we write $\rho = \sqrt{1-u^2}$. There is nothing
special about these choices, except that we use rather simple ones
in order to keep the resulting formulas simple enough; for the
same reason we will choose $B = \pi/2$.

The vector field $X_0$ in $U$ can then be completed to a vector
field $X$ in $\^M$ by the prescription $P^m = X_0 (A^m)$. This
yields
$$ P^1 = - B u \cos(|\a|) \ , \ \ P^2 = 0 \ , \ \ P^3 = B
u^2 \rho^{-1} \, \cos (|\a|) \ ; $$ note that on $\ga$ we have
$|\a| = B$, so that with our choice $B = \pi/2$ one gets simply
$P^m = 0$, $m=1,2,3$. The vector field $X$ is thus
$$
\begin{array}{rl} X \ =& \ - u \cos(|\a|) \pa_u \ + \ |\a|^{-1}
\sin(|\a|) \ \( \a^1 u \pa_v \, + \, \a^3 u \pa_w \, + \, \a^2 u
\pa_z \) \\ & \ \ - \ B u \cos(|\a|) \pa_1 + B u^2 \rho^{-1} \cos
(|\a|) \pa_3 \ .
\end{array} $$

It is immediate to check, see \eqref{eq:gainvar} above, that the
manifold $\^M_\ga$ corresponding to $\ga$ given by \eqref{eqsu2:A}
is invariant under $X$ (we recall this holds by construction). On
this manifold, $|\a| = B = \pi/2$ and $X$ reduces to
\beq\label{eqsu2:Xga} X_\ga \ = \ u^2 \, \pa_v \, + \, u \, \rho
 \, \pa_w \ . \eeq Finally, the projection of this to
a vector field in $M$ is simply $W = X_\ga$.

With this, we have completely described the upper row of the
diagram \eqref{diag:ZZ}. Let us now consider the lower one. First
of all we have to compute $X^{(2)}$, which turns out to be $$
\begin{array}{rl} X^{(2)} \ =& \ - \cos(|\a|) \( u \fracor{\pa}{\pa
u} + u_x \fracor{\pa}{\pa u_x} + u_{xx} \fracor{\pa}{\pa u_{xx}} \) \\
& \ + \ |\a|^{-1} \sin(|\a|) [ \a_1 u \fracor{\pa}{\pa v} + \a_1
u_x \fracor{\pa}{\pa v_x} + \a_1 u_{xx} \fracor{\pa}{\pa v_{xx}} ]
\\ & \ + \ |\a|^{-1} \sin(|\a|) [ \a_3 u \fracor{\pa}{\pa w} +
\a_3 u_x \fracor{\pa}{\pa w_x} + \a_3 u_{xx} \fracor{\pa}{\pa
w_{xx}} ]
\\ & \ + \ |\a|^{-1} \sin(|\a|) [ \a_2 u \fracor{\pa}{\pa z} +
\a_2 u_x \fracor{\pa}{\pa z_x} + \a_2 u_{xx} \fracor{\pa}{\pa z_{xx}} ] \\
 & \ - \ B u \cos(|\a|) \fracor{\pa}{\pa \a_1} + (B
u^2 \rho^{-1} ) \cos (|\a|) \fracor{\pa}{\pa \a_3} \ .
\end{array} $$
The restriction $X^{(2)}_\ga$ of this to $\^M^{(2)}_\ga$ is just
$$ \begin{array}{rl} X^{(2)}_\ga \ =& \ \ u^2 \fracor{\pa}{\pa v} +
\rho u \fracor{\pa}{\pa w} \ + \ u u_x \fracor{\pa}{\pa v_x} +
\rho u_x \fracor{\pa}{\pa w_x}
\\ &
\ + \ u u_{xx} \fracor{\pa}{\pa v_{xx}} + \rho u_{xx}
\fracor{\pa}{\pa w_{xx}} \ ;
\end{array} $$
the projection $Y$ of this to $M^{(2)}$ is of course $ Y =
X^{(2)}_\ga$.

Thus we get, using the notation \eqref{eq:exaY} for $Y$,
$$ \eta \ = \ - \, \pmatrix{ 0 \cr u^2 \cr \rho u
\cr 0 \cr} \ , \ \ \eta_x \ = \ - \, \pmatrix{0 \cr u u_x \cr \rho
u_x \cr 0 \cr} \ , \ \ \eta_{xx} \ = \ - \, \pmatrix{0 \cr u
u_{xx} \cr \rho u_{xx} \cr 0 \cr} \ . $$ We should then check that
these satisfy the relations \eqref{eq:exarecrel} with a suitable
$\La$; more precisely, in view of Theorem 1, with $\La = - (D_x
K_\ga) K_\ga^{-1}$ with $K_\ga$ the restriction of $K(\a)$ to the
section $\ga$.

Such a $\La$ can be computed using the general formulas given
above, or more simply from \eqref{eqsu2:K}. On $\ga$ we have $
|\a| = \sqrt{\a_1^2 + \a_2^2 + \a_3^2 } = B$; as for the matrix
$L$, on $\ga$ this is $$ L_\ga \ = \ B ( u L_1 + \rho L_3) \ = \ B
\ \pmatrix{0&u&\rho &0\cr -u&0&0&-\rho \cr -\rho &0&0&u\cr 0&\rho
&-u&0\cr} \ ; $$ hence on $\ga$ the matrix $J = |\a|^{-1} L$ is
just the square matrix appearing in the formula above. It follows
that $K_\ga$ and $K_\ga^{-1}$ are given by
$$ \begin{array}{l} K_\ga \ = \ \cos(B) \ I \ + \ \sin(B) \
\pmatrix{0&u&\rho &0\cr -u&0&0&-\rho \cr -\rho &0&0&u\cr 0&\rho
&-u&0\cr} \ , \\
K_\ga^{-1} \ = \ \cos(B) \ I \ - \ \sin(B) \ \pmatrix{0&u&\rho
&0\cr -u&0&0&-\rho \cr -\rho &0&0&u\cr 0&\rho &-u&0\cr} \ ,
\end{array} $$
as follows from \eqref{eqsu2:K}. These formulas are simplified by
the choice $B = \pi/2$, yielding
$$ K_\ga \ = \ \pmatrix{0&u&\rho &0\cr -u&0&0&-\rho \cr
-\rho &0&0&u\cr 0&\rho &-u&0\cr} \ , \ \ K_\ga^{-1} \ = \ - K_\ga
\ . $$

As for $\La$, this is immediately computed from the above and $\La
= - (D_x K_\ga) K_\ga^{-1}$, yielding for $B = \pi/2$ (we omit the
more complex formulas for the general case of arbitrary $B$)
$$ \La \ = \ \frac{1}{\sqrt{1-u^2}} \ \pmatrix{ 0&0&0&-u_x\cr
0&0&-u_x&0\cr 0&u_x&0&0\cr u_x&0&0&0\cr} \ . $$ One can then
easily check that the $\{\eta , \eta_x,\eta_{xx} \}$ given above
satisfy the prescribed relations, i.e. $ \eta_x = D_x \eta + \La
\eta$, $\eta_{xx} = D_x \eta_x + \La \eta_x$.

\subsection{Example 8.}

Let us consider a different examples for the same action of
$SU(2)$, now in the full $\R^4$ space. The section $\ga$ will now
correspond to \beq\label{eqsu2:A2} \a^1 = z \ , \ \ \a^2 = 2 \, z
\ , \ \ \a^3 = 5 \, z \ . \eeq We will moreover choose $\Theta =
(u,v,-z,w)$; thus the vector field to be gauged is a scaling in
the $(u,v)$ plane and a rotation in the $(w,z)$ one. The
corresponding gauged vector field in $U$ is $$ \begin{array}{rl}
X_0 =& \cos (\om) (u \pa_u + v \pa_v - z \pa_w + w \pa_z) \\ & + \
\om^{-1} \sin (\om) [ (\a_1 v + \a_2 w - \a_3 z) \pa_u - (\a_1 u +
\a_3 w + \a_2 z) \pa_v \\ & - (\a_3 u + \a_2 v - \a_1 w) \pa_w -
(\a_2 u - \a_3 v - \a_1 z) \pa_z ] \ . \end{array} $$ Here we have
written $\om = \sqrt{\a_1^2+\a_2^2+\a_3^2}$. With the usual
method, we get
$$ \begin{array}{l}
P^1 = w \cos(\om) - (\a_2 u - \a_3 v - \a_1 z) \om^{-1} \sin (\om) \ ,\\
P^2 = 2 [w \cos(\om) - (\a_2 u - \a_3 v - \a_1 z) \om^{-1} \sin (\om)] \ , \\
P^3 = 5 [w \cos(\om) - (\a_2 u - \a_3 v - \a_1 z) \om^{-1} \sin
(\om)] \ . \end{array} $$

We will thus consider the vector field in $\^M$ given, with again
$\pa_i := (\pa / \pa \a^i)$, by $$
\begin{array}{rl} X \ =& \ \cos (\om) [u \pa_u + v \pa_v - z \pa_w
+ w \pa_z + w (\pa_1 + 2 \pa_2 + 5 \pa_3)] \\ & + \ \om^{-1} \sin
(\om) [ (\a_1 v + \a_2 w - \a_3 z) \pa_u - (\a_1 u + \a_3 w + \a_2
z) \pa_v \\ & - (\a_3 u + \a_2 v - \a_1 w) \pa_w - (\a_2 u - \a_3
v - \a_1 z) \pa_z \\ & - (\a_2 u - \a_3 v - \a_1 z) (\pa_1 + 2
\pa_2 + 5 \pa_3)] \ . \end{array} $$

The manifold $\^M_\ga$ is invariant under this, and the
restriction of $X$ to $\^M_\ga$ is $$
\begin{array}{rl} X_\ga \ =& \ \cos (\Om) [u \pa_u + v \pa_v - z \pa_w
+ w \pa_z + w (\pa_1 + 2 \pa_2 + 5 \pa_3)] \\ & + \ \Om^{-1} \sin
(\Om) [ (v z + 2 w z - 5 z^2) \pa_u - ( u z + 5 w z + 2 z^2) \pa_v
\\ & - (5 u z + 2 v z -  w z) \pa_w - (2 u z - 5 v z -  z^2) \pa_z \\
 & - (2 u z - 5 v z -  z^2) (\pa_1 + 2 \pa_2 + 5 \pa_3)] \ ;
\end{array} $$ note that now $\om$ has been replaced by $\Om =
\sqrt{30} z$. The projection of this vector field to $M$ is $$
\begin{array}{rl} W \ =& \ \cos (\Om) [u \pa_u + v \pa_v - z \pa_w
+ w \pa_z + w (\pa_1 + 2 \pa_2 + 5 \pa_3)] \\ & + \ \Om^{-1} \sin
(\Om) [ (v z + 2 w z - 5 z^2) \pa_u - ( u z + 5 w z + 2 z^2) \pa_v
\\ & - (5 u z + 2 v z -  w z) \pa_w - (2 u z - 5 v z -  z^2) \pa_z
\ . \end{array} $$

Let us now consider prolongations; the second prolongation of $X$
in $\^M^{(2)}$ is computed by standard algebra, giving a rather
long formula which we omit.


The restriction $X^{(2)}_\ga$ of this to $\^M^{(2)}_\ga$ is
readily obtained via the substitution $\om \to \Om$ and those
given by \eqref{eqsu2:A2}; as for the projection $Y$ of
$X^{(2)}_\ga$ to $M^{(2)}$, in which we are mostly interested, in
the notation \eqref{eq:exaY}, this corresponds to
$$ \begin{array}{l}
\eta \ = \ \cos (\Om) \, \pmatrix{u\cr v\cr -z\cr w\cr} \ + \
(1/\sqrt{30}) \, \sin(\Om) \, \pmatrix{v + 2 w - 5 z \cr - (u + 5
w + 2 z) \cr - (5 u + 2 v - w) \cr
- 2 u + 5 v + z \cr} \ ; \\
{ } \\ \eta_x \ = \ \cos (\Om) \, \pmatrix{u_x\cr v_x\cr -z_x\cr
w_x\cr} \ + \ (1/\sqrt{30}) \, \sin(\Om) \, \pmatrix{v_x + 2 w_x -
5 z_x \cr - (u_x + 5 w_x + 2 z_x) \cr - (5 u_x + 2 v_x - w_x) \cr
- 2 u_x + 5 v_x + z_x \cr} \ ; \\
{ } \\
\eta_{xx} \ = \ \cos (\Om) \, \pmatrix{u_{xx}\cr v_{xx}\cr
-z_{xx}\cr w_{xx}\cr} \ + \ (1/\sqrt{30}) \, \sin(\Om) \,
\pmatrix{v_{xx} + 2 w_{xx} - 5 z_{xx} \cr - (u_{xx} + 5 w_{xx} + 2
z_{xx}) \cr - (5 u_{xx} + 2 v_{xx} - w_{xx}) \cr - 2 u_{xx} + 5
v_{xx} + z_{xx} \cr} \ . \end{array} $$

We should now check that relations \eqref{eq:exarecrel} are
satisfied for a suitable matrix $\La = - (D_x K_\ga) K_\ga^{-1}$.
In our case
$$ K_\ga \ = \  \cos (\Om) \, I \ + \ \frac{\sin(\Om)}{\sqrt{30}} \,
\pmatrix{0&1&5&2\cr -1&0&2&-5\cr -5&-2&0&1\cr -2&5&-1&0\cr} \ . $$
It follows that
$$ \La \ = \ - \, z_x \
\pmatrix{0&1&5&2\cr -1&0&2&-5\cr -5&-2&0&1\cr -2&5&-1&0\cr} \ . $$
One can easily check that this satisfies indeed the required
relations $$ \eta_x = D_x \eta + \La \eta \ , \ \ \eta_{xx} = D_x
\eta_x + \La \eta_x \ . $$

\subsection{Example 9.}

We will now turn -- still using the same real representation of
$G=SU(2)$ -- to examples dealing with Theorem 2.

Let us consider the vector field $Y$ corresponding, in the
notation \eqref{eq:exaY} and writing $\rho := \sqrt{v^2+z^2}$, to
\beq\label{eqsu2:eta3} \eta = u \[ \cos (\rho) \pmatrix{1\cr 0\cr
1\cr 0\cr} - \frac{\sin(\rho)}{\rho} \pmatrix{0\cr  (v-z)\cr 0\cr
(v+z)\cr} \] , \ \eta_x = \frac{u_x}{u} \eta , \ \eta_{xx} =
\frac{u_{xx}}{u} \eta \ . \eeq

This is a $\mu$-prolonged vector field, with $\mu = \La \d x$
identified by $$ \La \ = \ \rho^{-2} B_0 \ + \ (1/2) \rho^{-3}
\sin (2 \rho) \ B_1 \ + \ \rho^{-2} \sin^2 (\rho) \ B_2 \ , $$
where the matrices $B_i$ are given by $$
\begin{array}{l} B_0 = (v v_x + z z_x) \, \pmatrix{0& -v& 0&
-z\cr v & 0& -z & 0\cr 0& z & 0& -v \cr z & 0& v & 0\cr} \ , \\
B_1 = (v_x z - v z_x) \pmatrix{0& -z & 0 & v \cr z
& 0& v & 0\cr 0& - v & 0& -z \cr - v & 0& z & 0\cr} \ , \\
B_2 = (v_x z - v z_x) \pmatrix{0& 0& 1& 0\cr 0& 0& 0& -1\cr -1& 0&
0& 0\cr 0& 1& 0& 0\cr} \ . \end{array} $$

This $\La$ is in the general form given above provided $\ga$ is
identified by $$ \a^1 = v \ , \ \a^2 = z \ , \ \a^3 = 0 \ . $$
Having identified $\ga$ allows in turn to identify the gauging
matrix $K_\ga$ as
$$ K_\ga = \cos (\rho) \, I \ + \ \rho^{-1} \sin (\rho) \pmatrix{
0& v & 0& z \cr -v & 0& z & 0\cr 0& -z & 0& v \cr -z & 0& -v &
0\cr} \ ; $$ hence the vector field $X_\ga^{(2)}$ is also
determined and -- with the procedure described in Section
\ref{sect:main} -- the full $X^{(2)}$ is readily obtained as well
(we omit the involved explicit formula); this is the standard
prolongation of
$$ \begin{array}{rl} X \ =& \ \cos(\om) \(  u \pa_u + u \pa_w \)
\ - \ \om^{-1} \sin(\om ) u \[ \a_3 \pa_u + (\a_2 - \a_1) \pa_v -
\a_3 \pa_w \right. \\ & \left. + (\a_2 - \a_1) \pa_z + \pa_1 -
\pa_3 \] \ . \end{array} $$

This leaves $\ga$ invariant, and its components along $U$
correspond to a gauged vector field, being of the form $\phi^a =
[K(\a)]^a_{\ b} \Theta^b$ with $ \Theta = (1,0,0,0)$.

\subsection{Example 10.}

We will now consider the vector field $Y$ in $M^{(2)}$ given
(writing $\Om = \sqrt{30} z$) by
$$ Y \ = \ \cos(\Om ) \( \pa_u + \pa_w \) \ + \
\frac{\sin(\Om)}{\sqrt{30}} \( 5 \pa_u + \pa_v - 5 \pa_w - 3 \pa_z
\) \ ; $$ that is, with the notation \eqref{eq:exaY} we have
$$ \eta = \cos(\Om) \, \pmatrix{1\cr 0\cr 1\cr 0\cr} +
\frac{z \sin(\Om)}{\Om} \, \pmatrix{5\cr 1\cr -5\cr -3\cr} \ , \
\eta_x = \pmatrix{0\cr 0\cr 0\cr 0\cr} \ , \ \eta_{xx} =
\pmatrix{0\cr 0\cr 0\cr 0\cr} \ . $$

This is a $\mu$-prolonged vector field, with
$$ \La \ = \ z_x \ \pmatrix{0& -1& -5& -2\cr 1& 0& -2& 5\cr
5& 2& 0& -1\cr 2& -5& 1& 0\cr} \ ; $$ this in turn corresponds,
see our general formulas above, to $\ga$ identified by
$$ \a_1 = z \ , \ \a_2 = 2 z \ , \ \a_3 = 5 z \ . $$
We can in this way identify $X^{(2)}_\ga$ and $X^{(2)}$; the
latter turns out to be
$$ \begin{array}{rl}
X^{(2)} \ =& \ \cos (\om) \ \( \pa_u + \pa_v \) \ + \ [\sin(\om) /
\om ] \ \( \a_3 \pa_u +(\a_2 - \a_1) \pa_v
 - \a_3 \pa_w  \right. \\
 & \left. \ \ - (\a_1 + \a_2) \pa_z
 \ - \ (\a_1+\a_2) (\pa_1 + 2 \pa_2 +5 \pa_3 ) \) \ .
 \end{array} $$
The vector field $X$ just coincides with $X^{(2)}$; its components
along $U$, given by $\phi = \cos(\om) (1,1,0,0) + \om^{-1}
\sin(\om) (\a_3, (\a_2 - \a_1), - \a_3 , - (\a_1 + \a_2) )$,
correspond to a gauged vector field obtained for the choice
$\Theta = (1,0,1,0)$. The gauging matrix $K(\a)$ can be derived
either by this or noticing that the $\La$ given above corresponds
to
$$ K_\ga \ = \ \cos (\Om) \, I \ ; $$
with the usual prescription this yields
$$ K (\a) \ = \ \cos (\om ) \ I \ + \ \frac{\sin(\om)}{\om} \
\pmatrix{0& a_1& a_3& a_2\cr -a_1& 0& a_2& -a_3\cr -a_3& -a_2& 0&
a_1\cr -a_2& a_3& -a_1& 0\cr} \ . $$ The same result is obtained
comparing $X$ and the $\Theta$ given above.

\section{Examples III. Non abelian groups: SO(3)}

In this section we will consider the group $G=SO(3)$ acting in
$R^3$ by its natural representation; once again we will consider
very simple vector fields and section. Coordinates in $U$ will be
denoted by $(u,v,w)$. Example 11 deals with Theorem 1, while
Example 12 with Theorem 2.

\subsection{SO(3) algebra and group action; lambda matrices}

We will consider generators $L_i = T (\ell_i)$, $$ L_1 =
\pmatrix{0&0&0\cr 0&0&-1\cr 0&1&0\cr} , L_2  = \pmatrix{0&0&1\cr
0&0&0\cr -1&0&0\cr} , L_3 = \pmatrix{0&-1&0\cr 1&0&0\cr 0&0&0\cr}
\ . $$

The group element $g$ corresponding to $g = \exp(\ell)$ for $\ell$
a generic element of the algebra is readily computed. Consider a
generic matrix $ L = \a^i L_i$; this is written explicitly as
$$ L \ = \ \pmatrix{0&-\a_3&\a_2\cr
\a_3&0&-\a_1\cr -\a_2&\a_1&0\cr} \ . $$ Some easy computations
show that higher powers of $L$ satisfy ($k \ge 0$)
$$ L^{2k+1} \ = \ (-1)^{k} \om^{2k} \, L \ , \ \ L^{2 (k+1)} \ = \
(-1)^{k} \om^{2k} \, L^2 \ , $$ where we used $$ \om =
\sqrt{\a_1^2 + \a_2^2 + \a_3^2} \ . $$ Using the Taylor expansions
of trigonometric functions, it follows that $K(\a) = \exp(L)$ and
its inverse can be written as $$ \begin{array}{l} K(\a) \ = \ I \
+ \ \om^{-1} \sin(\om) \, L \ + \ \om^{-2} [1 - \cos(\om)] \, L^2
\\ K^{-1} (\a) \ = \ I \ - \ \om^{-1} \sin(\om) \, L \ + \
\om^{-2} [1 - \cos(\om)] \, L^2 \ . \end{array} $$ We will not
give the general expression of matrices $\La = - (D_x K_\ga)
K_\ga^{-1}$ corresponding to $K$, $K^{-1}$ given above, as the
formula -- which can be readily derived with the help of a
symbolic manipulation program -- is quite involved; its general
shape is $ \La = M_0 + \sin (\om )  M_1 + \cos (\om ) M_2$, where
the $M_i$ are three-dimensional matrices.\footnote{We stress this
$\La$ depends on the three arbitrary smooth functions $A_i
(x,u,v,w,z)$ and their derivatives. Thus if we want to identify a
given three-dimensional matrix with $\La$, this yields a system of
PDEs for the three functions $A_i$.}

The general form of gauged vector fields in $U$ is easily obtained
applying $K(\a)$ on $\Theta = (\th^1,\th^2,\th^3)$; conversely,
given a gauged vector field with components $\Phi =
(\phi^1,\phi^2,\phi^3)$, the corresponding $\Theta$ is given by
$\Theta = [K^{-1} (\a)] \Phi$.

\subsection{Example 11.}

We will consider $\R^3$ with cartesian coordinates $(u,v,w)$ and
$U \ss \R^3$ defined by $u < 1$. The vector field $X_0$ in $U$
will be
$$ \begin{array}{rl} X_0 \ =& \ u \om^{-2} \ \[ \( \a_1^2 + (\a_2^2+\a_3^2) \cos
(\om ) \) \pa_u \ + \ \( \a_1 \a_2 (1 - \cos (\om ) ) + \a_3 \om
\sin (\om) \) \pa_v \right. \\ & \left. \ + \( \a_1 \a_3 (1 -
\cos(\om)) - \a_2 \om \sin (\om ) \) \pa_w \] \ , \end{array} $$
where as usual $\om = \sqrt{\a_1^2 + \a_2^2 + \a_3^2}$. This is a
gauged vector field, corresponding to the choice $\Theta =
(u,0,0)$.

We choose the section $\ga$ identified by
$$ \a^1 = \b u \ , \ \a^2 = 0 \ , \
\a^3 =  \b \sqrt{1 - u^2} \ , $$ where $\b$ is an arbitrary real
constant; formulas are simpler with the choice $\b= \pi/2$. The
vector field $X_0$ is completed to a vector field $X = X_0 + P^m
\pa_m$ in $\^M$, leaving $\ga$ invariant, with the choice $ P^m =
X_0 (A^m)$.

Using the notation $\rho = \sqrt{1-u^2}$, the restriction of $X$
to $\gamma$ is
$$ \begin{array}{rl} X_\ga \ =& \ \( u + u^3 (1 - \cos \b ) \) \pa_u + \( u \rho
\sin \b \) \pa_v + \( 2 u^2 \rho \sin^2 (\b/2) \) \pa_w \\ & \ + \
\b u \( u^2 + (1 - u^2) \cos \b \) \pa_1 - \b u^2 \rho^{-1} \( u^2
+ (1-u^2) \cos \b \) \pa_3 \ ; \end{array} $$ the projection of
this to $M$ is easily computed to be
$$ W \ = \ \( u + u^3 (1 - \cos \b ) \) \pa_u + \( u \rho
\sin \b \) \pa_v + \( 2 u^2 \rho \sin^2 (\b/2) \) \pa_w \ . $$

We can compute with standard procedure the second prolongation of
$X$, restrict it to $\^M^{(2)}_\ga$ and project to $M^{(2)}$. The
final result is in the form ((28)) with
$$ \eta = u \pmatrix{u^2 + (1-u^2) \cos \b \cr \rho \sin \b \cr 2
\rho u \sin^2 (\b/2) \cr} \ , \ \eta_x = \frac{u_x}{u} \ \eta \ ,
\ \eta_{xx} = \frac{u_{xx}}{u} \, \eta \ . $$ These should be
checked to satisfy the relations ((29)) with $\La$ given by
Theorem 1. In the present case,
$$ K_\ga = \pmatrix{u^2 + (1-u^2) \cos \b & - \rho \sin \b & \rho
u (1 - \cos \b) \cr \rho \sin \b & \cos \b & - u \sin \b \cr \rho
u (1 - \cos \b) & u \sin \b & (1-u^2) + u^2 \cos \b \cr} \ ; $$
$$ K_\ga^{-1} \ = \ \pmatrix{u^2 + (1-u^2) \cos \b & \rho \sin \b & \rho
u (1 - \cos \b) \cr - \rho \sin \b & \cos \b & u \sin \b \cr \rho
u (1 - \cos \b) & - u \sin \b & (1-u^2) + u^2 \cos \b \cr} \ . $$

With standard computations, we get first $(D_x K_\ga)$ and then
$$ \La = - (D_x K_\ga) \, K_\ga^{-1} \ = \ \frac{u_x}{\rho} \
\pmatrix{0& - u & - 1 \cr u  \sin \b & 0 & \rho  \sin \b \cr 2
\sin^2 (\b/2) & - \rho  \sin \b & 0 \cr} \ . $$ It is easily
checked that indeed ((29)) are satisfied.

In  the simple case $\b = \pi/2$, we are reduced to
$$
\eta = u \pmatrix{u^2\cr\rho\cr\rho u\cr}  , \, \eta_x =
(\frac{u_x}{u} \eta , \, \eta_{xx} = \frac{u_{xx}}{u} \eta ; \ \La
= \frac{u_x}{\rho} \pmatrix{0&- u & - 1 \cr u & 0 & \rho \cr 1 & -
\rho & 0 \cr} \ .
$$

\subsection{Example 12}

Let us consider the vector field $Y$ given, in the notation ((28))
and using the conventions set in the previous Example, by
$$ \eta = u \, \pmatrix{- \rho w \cr 0 \cr u w \cr} \ , \
\eta_x = u_x \, \pmatrix{- \rho w \cr 0 \cr u w \cr} \ , \
\eta_{xx} = u_{xx} \, \pmatrix{- \rho w \cr 0 \cr u w \cr} \ . $$
These satisfy ((29)) with
$$ \La \ = \ \rho \ \pmatrix{0& - u u_x & - u_x \cr u u_x & 0 &
\rho u_x \cr u_x & - \rho u_x & 0 \cr} \ . $$

In order to associate this with a gauging matrix $K_\ga$, one can
either proceed by massive computations using the general form of
$\La$ in terms of the functions $A^i (x,u,v,w)$, or observe that
only the $u$ variable appears in $\La$, and proceed by trial and
error to determine
$$ K_\ga \ = \ \pmatrix{u^2 & - \rho & \rho u \cr \rho & 0 & - u
\cr \rho u & u & 1- u^2 \cr } \ {\rm with} \ K_\ga^{-1} \ = \
\pmatrix{u^2 & \rho & \rho u \cr - \rho & 0 & u \cr \rho u & - u &
1- u^2 \cr } \ . $$

At this point it suffices to use $ K_\ga^{-1}$ to transform the
$(\eta,\eta_x,\eta_{xx})$ into $\th = K_\ga^{-1} \eta$ etc.; we
get
$$ \th = \pmatrix{0\cr u^2 w  + u (1-u^2) w \cr 0 \cr} \ , \
\th_x = u_x \, \th \ , \ \th_{xx} = u_{xx} \, \th \ . $$ It is
easy to check that $\th_x = D_x \th$ and $\th_{xx} = D_x \th_x$,
i.e. the vector field
$$ \th^a \fracor{\pa}{\pa u^a} \ + \ \th^a_x \fracor{\pa}{\pa u^a_x} \
+ \ \th^a_{xx} \fracor{\pa}{\pa u^a_{xx}} $$ is the standard
second prolongation of
$$ X_0 \ = \ \( u^2 w  + u (1-u^2) w \) \ \fracor{\pa}{\pa v} \ . $$

In order to complete this to a vector field in $\^M$, we compare
$K_\ga$ and the general expression for $K (\a )$, and observe that
we obtain such a $K_\ga$ by choosing $\ga$ identified by
$$ \a^1 = \fracor{\pi}{2} u \ , \ \a^2 = 0 \ , \
\a^3 = \fracor{\pi}{2} \sqrt{1-u^2} \ . $$ Applying $X_0$ on the
functions $X^i (x,u,v,w)$ defined by these relations, we get $P^1
= \fracor{\pi}{2} u$, $P^2 =0$, $P^3 = \fracor{\pi}{2}
\sqrt{1-u^2}$.

\vfill\eject

\section{Discussion}
\label{sec:discussion}

We have thus shown that $\mu$-prolongation can be understood as a
standard prolongation in the gauge jet bundle $\^M^{(k)}$,
restricted to a section of the bundle and then projected to the
$M^{(k)}$ bundle.

With this point of view on $\mu$-prolongations -- and on the
discussion in the present work -- there are some observations to
be made and points to be stressed; some of these suggest in turn
further developments.
\bigskip

{\bf (1)} We were able to conduct our discussion within the
framework of gauged vector fields; acting by a gauge
transformation $\ga \in \Ga (P_G)$ in $\^M^{(k)}$, the bundle $J^k
M$, which in this context should be seen as a submanifold of
$\^M^{(k)}$, is mapped into $\^M_\ga^{(k)}$, and prolonged vector
fields are mapped into vector fields which, if projected back to
$J^k M$, appear as $\mu$-prolonged ones. By acting with the
inverse transformation, the $\mu$-prolonged vector field can be
transformed back into a standard prolonged vector field.

{\bf (2)} Roughly speaking, this shows that the possibility to
apply $\mu$-prolongations and $\mu$-symmetries of differential
equations with the same effectiveness as standard prolongations
and symmetries can be understood as a consequence of the fact that
the differential equations under consideration are written in
terms of ordinary rather than covariant derivatives; this in turn
makes that the equations are set (with the language employed in
this note) in the ordinary bundle $J^k M$ rather than in the
augmented bundle $J^k \^M$.

{\bf (3)} In this way, a gauge transformation maps the
differential equation $\Delta$ under study into a different
equation $\wt{\Delta}$, which can admit (ordinary) symmetries not
admitted by the original equation as ordinary symmetries, albeit
they are admitted as $\mu$-symmetries. Thus one could say that the
approach devised by Muriel and Romero is somehow opposite to the
one which is standard in field theory: rather than promoting PDEs
to covariant equations (that is, write them in terms of covariant
derivatives), one keeps non-covariant equations and uses gauge
transformation to maps them to different equations. This procedure
represents an advantage if the gauge orbit of the considered
equation $\Delta$ contains an equation $\wt{\Delta}$ with a higher
symmetry.

{\bf (4)} This point of view also suggests an obvious approach to
apply $\mu$-sym\-me\-tri\-es for the reduction of general systems
of differential equations by differential invariants (so far a
procedure for this is known only when the $\La_i$ satisfy some
additional conditions \cite{Cicrho,MuTMP,MuVigo}; it is not known
if these are only sufficient or also necessary). That is, work in
the whole $J^k \^M$, where differential invariants of higher
orders can be obtained from lower order ones by the familiar
recursive procedure, and then restrict to the relevant sub-bundle
$J^k \^M_\ga$. Such an approach is however too simple to work; the
reason is that, due to the term $P^m (\pa / \pa \a^m)$, the
standard recursive procedure \cite{Olv1} does not in general give
new invariants\footnote{This is related to the fact we now have,
for general vector fields and with standard notation,
$[X^{(k)},D_x] = - (D_x \xi) D_x - (D_x P^m) (\pa / \pa \a^m)$
rather than just $[X^{(k)},D_x] = - (D_x \xi) D_x$.} and should be
modified accordingly. This will be considered in a separate
contribution.

{\bf (5)} Note also, making free use of the notation introduced in
the proof to Theorem 1, that the reduction from vector fields in
$\^M$ to vector fields in $\^M_\ga$ was natural for the diagram
(\ref{diag:Z00}) provided the relation $\vth^m = (\pa A^m / \pa
u^a) \vphi^a$ was satisfied, and this independently of the
condition (\ref{eq:sep}) for $X$. On the other hand, the operator
${\tt \^P_\ga}$ and hence ${\tt P_\ga}$ (which then turned out to
be the $\mu$-prolongation operator with suitable $\mu$) was at
first defined as the operator making the left-hand side of
(\ref{diag:Z00}) commutative. Thus, it may also be defined
independently of (\ref{eq:sep}). In principles, this could give a
generalization of the $\mu$-prolongation operation. This problem
will also be considered elsewhere.

{\bf (6)} It should also be mentioned that here we worked with
sections of $\gaub$ as basic objects, corresponding to the
requirement $g = g(x,u)$. One could start from sections of $J^s
\gaub$, corresponding to gauge transformations with $g =
g(x,u^{(s)})$, i.e. depending not only on the space-time point $x$
and on the values of fields $u^a$ at $x$, but also on the values
of field derivatives up to order $s$ at $x$.

{\bf (7)} The point (3) of the present discussion suggests that in
order to deal with $\mu$-symmetries of differential equations set
in terms of standard partial derivatives, it might be convenient
to rewrite them in terms of covariant derivatives (that is, write
partial derivatives $u_i$ as $u^a_i = \grad_i u^a - (\La_i)^a_{\
b} u^b$, and the like for higher derivatives), extending them to
covariant equations in the $\^M$ space -- allowing of course also
changes in the $\La_i$ matrices and the reference $\gaub$ section
-- in order to take full advantage of the gauge formalism.

{\bf (8)} As for physical relevance, the analysis here (and that
in \cite{Gframe}, respectively) considered above show that with
the formalism of $\mu$-symmetries one can be able to detect
symmetries and hence conserved quantities even if working in a
non-convenient gauge (respectively, reference frame). This opens
the interesting possibility of applying symmetry analysis and
Noether's theorem (see \cite{CGnoet} in this respect) also when
working in a gauge in which the equations are not manifestly
symmetric.

{\bf (9)} The point of view embodied in this work, based on the
gauge bundle, has several points of contact with those explored in
recent works by other authors: P. Morando considered a gauging of
the exterior derivative and showed how this leads to
$\mu$-symmetries \cite{Mor}; D. Catalano-Ferraioli considered
auxiliary variables in the context of the theory of {\it
coverings}, showing intriguing relations between (local)
$\la$-symmetries and nonlocal standard symmetries \cite{Cat}; see
also \cite{MuRom07}. This relations are also used in connection to
{\it solvable structures } in \cite{CM08}. Relation with this
approach is briefly discussed in the Appendix below.

{\bf (10)} Finally, we recall that -- as discussed in \cite{CGM}
-- the case of a scalar ODE, originally considered by Muriel and
Romero \cite{MuRom1}, is degenerate in several ways. These
degenerations hide the rich geometrical structure displayed in the
general case of system of PDEs, i.e. passing from $\la$ to
$\mu$-prolongations, and make that the case considered at first is
actually the most difficult one. Needless to say, this makes the
work by Muriel and Romero even more remarkable.

\vfill\eject

\section*{Appendix. Gauge variables as auxiliary variables}

When we fix a gauge, i.e. set $\a^m = A^m (x,u)$, we are actually
prescribing a correspondence between functions $u^a = f^a (x)$ and
expressions of $\a$ as functions of the $x$ themselves, via $\a^m
= F^m (x) := A^m (x, f(x))$. In this sense, gauge fixing is
equivalent to a constraint relating the $x,u$ and the $\a$ seen as
auxiliary dependent variables. This description of gauge variables
as auxiliary dependent variables is reminiscent of the approach to
$\lambda$-symmetries via the formalism of coverings \cite{Cat},
and we would thus like in this Appendix to discuss how our
construction can be modified to take this point of view into
account.

Let us consider ``fully augmented'' bundles $\M = M \times \G$, in
which the gauge variables $\a$ should be seen as new dependent
variables. Thus, in the corresponding jet bundles $J^k \M =
\M^{(k)}$ will also appear variables $\a^m_J$ corresponding to
derivatives of the $\a$, and the contact structure in $\M^{(k)}$
will also include forms $\Xi^m_J = \d \a^m_J - \a^m_{J,i} \d x^i$.
The total derivatives operators $\D_i$ in $\M^{(k)}$ will thus be
$$ \D_i \ = \ \frac{\pa}{\pa x^i} \ + \ u^a_{J,i} \,
\frac{\pa}{\pa u^a_J} \ + \ \a^m_{J,i} \, \frac{\pa}{\pa \a^m_J} \
= \ D_i \ + \ Z_i \ ; $$ here we have of course defined $Z_i =
\a^m_{J,i} (\pa / \pa \a^m_J)$ and denoted again by $D_i$ the
usual total derivative operator in $M^{(k)}$.

Now the gauge fixing operator $\de_\ga$ corresponds to introducing
the constraint $\ga$ given by $\a^m - A^m (x,u) = 0$. For any
function $F (x,u,\a)$ we have immediately $$ \de_\ga \, [ D_i F] \
= \ \[ (\pa F /\pa x^i) \]_\ga \ + \ u^a_i \, \[ /\pa F/\pa u^a)
\]_\ga \ . $$ On the other hand, $$ D_i \, [\de_\ga F] \ = \ \[ (\pa
F/\pa x^i) \]_\ga \ + \ u^a_i \, \[ (\pa F/\pa u^a) \]_\ga \ + \
\[D_i A^m (x,u)\] \, \[ (\pa F/\pa \a^m) \]_\ga \ . $$ Comparing these
two expressions we get $D_i [\de_\ga F] = \de_\ga [ (D_i + Z_i
)(F)]$, which can also be written as $$ \de_\ga \ ( \D_i \, F ) \
= \ D_i \ (\de_\ga \, F) \ , \eqno(A.1)
$$ or equivalently
$$ \de_\ga \ (D_i \, F) \ = \ D_i \ (\de_\ga \, F ) \ -
\ \de_\ga \ (Z_i F) \ . \eqno(A.2) $$ The relation (A.1) describes
how total differential operators should be modified if applied
before or after gauge fixing.

In particular, if we apply (A.2) on $Q^a (x,u,\a,u_x) = [K
(\a)]^a_{\ b} \Q0^b (x,u,u_x)$, we have $Z_i Q^a = (Z_i K^a_{\ b})
\Q0^b = [(Z_i K) K^{-1}]^a_{\ b} Q^b$. It is easy to see that $$
\de_\ga [ (Z_i K) K^{-1}] = (D_i K_\ga ) K_\ga^{-1} \ := \
R_i^{(\ga)} , $$ where as before we wrote $K_\ga = \de_\ga (K)$
and $R_i^{(\ga)}$ is also defined as above.

That is, if we first act with $D_i$ and then fix the gauge we
obtain the same result as by first fixing the gauge and then
acting with $\nabla_i^{(\ga)} = D_i - R_i^{(\ga)}$. We can then
identify the twisted prolongation obtained by acting with
operators $\nabla_i^{(\ga)}$ (rather than $D_i$) with a
$\mu$-prolongation, proceeding in the same way -- and with the
same $\mu$ -- as in the main text.


\begin{thebibliography}{99}

\bibitem{Ble} D. Bleecker, {\it Gauge theory and variational principles},
Addison-Wesley, Reading 1981; reprinted by Dover, Mineola 2005

\bibitem{Cat} D. Catalano Ferraioli, ``Nonlocal aspects of
$\la$-symmetries and ODEs reduction'', {\it J. Phys. A} {\bf 40}
(2007) 5479-5489

\bibitem{CM08} D. Catalano Ferraioli and P. Morando, ``Local and nonlocal
solvable structures in ODEs reduction'', {\it J. Phys. A} {\bf 42}
(2009) 035210

\bibitem{CCS} S.S. Chern, W.H. Chen and K.S. Lam, {\it Lectures on
differential geometry}, World Scientific, Singapore 1999

\bibitem{Cic04} G. Cicogna, ``Weak symmetries and adapted variables
for differential equations'', {\it Int. J. Geom. Meths. Mod.
Phys.} {\bf 1} (2004), 23-31

\bibitem{Cicrho} G. Cicogna,
``Reduction of systems of first-order differential equations via
$\Lambda$-symmetries'', {\it Phys. Lett. A} {\bf 372} (2008),
3672-3677

\bibitem{CGnoet} G. Cicogna and G. Gaeta, ``Noether theorem for
$\mu$-symmetries'', {\it J. Phys. A} {\bf 40} (2007), 11899-11921

\bibitem{CGM} G. Cicogna, G. Gaeta and P. Morando, ``On the
relation between standard and $\mu$-symmetries for PDEs'', {\it J.
Phys. A} {\bf 37} (2004), 9467-9486

\bibitem{EGH} T. Eguchi, P.B. Gilkey and A.J. Hanson, ``Gravitation, gauge
theories and differential geometry'', {\it Phys. Rep.} {\bf 66}
(1980), 213-393

\bibitem{Gbook} G. Gaeta, {\it Nonlinear symmetries and nonlinear equations},
Kluwer, Dordrecht 1994

\bibitem{Gframe} G. Gaeta, ``Smooth changes of frame and prolongations
of vector fields'', {\it Int. J. Geom. Meths. Mod. Phys.} {\bf 4}
(2007), 807-827

\bibitem{GM} G. Gaeta and P. Morando, ``On the geometry of
lambda-symmetries and PDEs reduction'', {\it J. Phys. A} {\bf 37}
(2004), 6955-6975

\bibitem{GoS} M. G\"ockeler and T. Sch\"ucker, {\it Differential geometry, gauge
theories, and gravity}, Cambridge University Press, Cambridge 1987

\bibitem{Ish} C.J. Isham, {\it Modern differential geometry for
physicists}, World Scientific, Singapore 1999 (2nd edition)

\bibitem{Vin} I.S. Krasil'schik and A.M. Vinogradov (eds.), {\it Symmetries and
conservation laws for differential equations of Mathematical
Physics}, A.M.S., Providence 1999

\bibitem{Mor} P. Morando, ``Deformation of Lie derivative and
$\mu$-symmetries'', {\it J. Phys. A} {\bf 40} (2007), 11547-11559

\bibitem{MuRom1} C. Muriel and J.L. Romero, ``New method of reduction for
ordinary differential equations'', {\it IMA Journal of Applied
mathematics} {\bf 66} (2001), 111-125

\bibitem{MuTMP} C. Muriel and J.L. Romero, ``Prolongations of vector
fields and the property of the existence of invariants obtained by
differentiation'', {\it Theor. Math. Phys.} {\bf 133} (2002),
1565-1575

\bibitem{MuVigo} C. Muriel and J.L. Romero, ``$C^\infty$ symmetries and
integrability of ordinary differential equations'', in {\it
Proceedings of the I colloquium on Lie theory and applications}
(I. Bajo and E. Sanmartin eds.), Publicaci\'ons da Universidade de
Vigo, Vigo 2002

\bibitem{MuRom07} C. Muriel and J.L. Romero, ``$C^\infty$-symmetries
and nonlocal symmetries of exponential type'', {\it IMA J. Appl.
Math.} {\bf 72} (2007), 191-205

\bibitem{MRO} C. Muriel, J.L. Romero and P.J. Olver, ``Variational
$C^\infty$ symmetries and Euler-Lagrange equations'', {\it J.
Diff. Eqs.} {\bf 222} (2006) 164-184

\bibitem{Nak} M. Nakahara, {\it Geometry, Topology and
Physics}, IOP, Bristol 1990

\bibitem{NaS} C. Nash and S. Sen, {\it Topology and geometry for physicists},
Academic Press, London 1983

\bibitem{Olv1} P.J. Olver, {\it Application of Lie groups to differential
equations}, Springer, Berlin 1986

\bibitem{PuS} E. Pucci and G. Saccomandi, ``On the reduction methods for
ordinary differential equations'', {\it J. Phys. A} {\bf 35}
(2002), 6145-6155

\bibitem{Ste} H. Stephani, {\it Differential equations. Their solution using
symmetries}, Cambridge University Press, Cambridge 1989

\bibitem{Str} S. Sternberg, {\it Lectures on differential geometry},
Chelsea, New York 1983

\end{thebibliography}
\end{document}